  \documentclass{aa}
   \usepackage{graphicx}
   
\begin{document}
   \title{The 3-D shaping of 
NGC 6741: a massive, fast-evolving Planetary Nebula at the recombination--reionization edge.
 \thanks{Based on observations made with: ESO Telescopes at the La Silla 
Observatories (program ID 65.I-0524), 
and the NASA/ESA Hubble Space Telescope, 
obtained from the data archive at the Space Telescope Institute. 
Observing programs: GO 7501 and GO 8773 (P.I. Arsen Hajian). 
STScI is operated by the association of Universities for Research in   
Astronomy, Inc. under the NASA contract  NAS 5-26555.  We extensively 
apply the photo--ionization code CLOUDY, developed at the Institute of
Astronomy of the Cambridge University (Ferland et al. 1998).}
 \author{ F. Sabbadin  \inst{1} \and S. Benetti \inst{1} \and E. Cappellaro \inst{2} \and R. Ragazzoni \inst{3} \and M. Turatto \inst{1} }
   \offprints{F. Sabbadin, sabbadin@pd.astro.it}
   \institute{INAF - Osservatorio Astronomico di Padova, vicolo dell'Osservatorio 5, I-35122 Padova, Italy \and 
   INAF - Osservatorio Astronomico di Capodimonte, via Moiariello 11, I-80131 Napoli, Italy \and INAF - Osservatorio 
   Astrofisico di Arcetri, Largo E. Fermi 5, I-50125, Italy 
}}
   \date{Received November 29, 2004; accepted March 9, 2010}
   \abstract{We infer the gas kinematics, diagnostics and ionic radial profiles, distance and central
   star parameters, nebular photo-ionization model, spatial 
structure and evolutionary phase of the Planetary Nebula NGC 6741 by means of long-slit ESO NTT+EMMI high-resolution 
spectra at nine position angles, reduced and analysed according to the tomographic and 3-D methodologies developed at the Astronomical Observatory 
of Padua (Italy). 
\\
NGC 6741 (distance$\simeq$2.0 kpc, age$\simeq$1400 yr, 
ionized mass M$_{\rm ion}\simeq$0.06 M$_\odot$) is a dense (electron density up to 12\,000 cm$^{-3}$), high-excitation, almost-prolate ellipsoid 
(0.036 pc x 0.020 pc x 0.018 pc, major, intermediate and minor semi-axes, respectively), surrounded by a sharp low-excitation skin (the ionization 
front), and embedded into a spherical (radius$\simeq$0.080 pc), almost-neutral, high-density ($n{\rm (H\,I)}\simeq$7$\times$10$^3$ atoms 
cm$^{-3}$) halo 
containing a large fraction of the nebular mass (M$_{\rm halo}\ge$0.20 M$_\odot$). 
The kinematics, physical conditions and ionic structure indicate that NGC 6741 is in a deep recombination 
phase, started about 200 years ago, and caused by the quick luminosity drop of the massive (M$_*$=0.66-0.68 
M$_\odot$), hot (logT$_*$$\simeq$5.23) and faint (log L$_*$/L$_\odot$$\simeq$2.75) post--AGB star, which has exhausted the hydrogen-shell nuclear 
burning and is moving along the white dwarf cooling sequence. The general expansion law of the ionized gas in NGC 6741, 
$V_{\rm exp}$(km s$^{-1}$)=13$\times$R$\arcsec$, fails in the innermost, highest-excitation 
layers, which move slower than expected. The observed deceleration is ascribable to the luminosity drop of the central star (the decreasing 
pressure of the hot-bubble no more balances the pressure of the ionized gas), and appears in striking contrast to recent reports inferring that 
acceleration is a common property of the Planetary Nebulae innermost layers. A detailed comparative analysis proves that the latter (i. e. 
``U''-shaped expansion velocity field) is a spurious, 
incorrect result due to a 
combination of: (a) simplistic assumption (spherical shell hypothesis for the nebula), (b) unfit reduction method (emission profiles integrated 
along the slit), and (c) inappropriate diagnostic choice ($\lambda$4686 $\rm\AA\/$ of He II, i. e. a thirteen fine-structure components 
recombination line). Some general implications on the shaping mechanisms of Planetary Nebulae are discussed.
   }
   
   \titlerunning{The recombining Planetary Nebula NGC 6741}
   
   \maketitle
%
\section{Introduction} 
\begin{figure} \centering
\includegraphics[width=9cm]{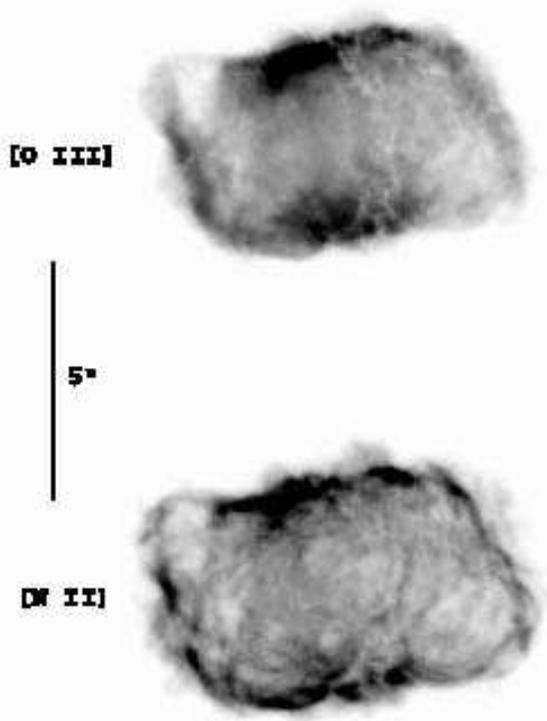} 
\caption{Apparent morphology of NGC 6741 in the medium-excitation line $\lambda$5007 $\AA\/$ of [O III] (upper panel), and  in the 
low-excitation line $\lambda$6584 $\rm \AA\/$ of [N II] (lower panel). North is up and East to the left. HST-WFPC2 frames; programme 
GO 8773, P.I. Arsen Hajian.}  
\end{figure}

The sentence by the late Professor Lawrence H. Aller (1994): ``A nebula is a three-dimensional structure for which we obtain a 
two-dimensional projection'' fully synthesizes the many, so far unsolved, observational limitations and interpretation problems connected to the 
Planetary Nebula (PN) research, leading to: (a) rough spatio-kinematical reconstruction, (b) unrealistic assumptions for the gas 
parameters (in particular, electron 
temperature ($T_{\rm e}$) and electron density ($N_{\rm e}$) constant all across the nebula), and (c) proliferation of kinematical, 
physical and evolutional models, frequently based on the mere nebular morphology (i. e. appearance).   
\begin{figure} \centering
\includegraphics[width=9cm]{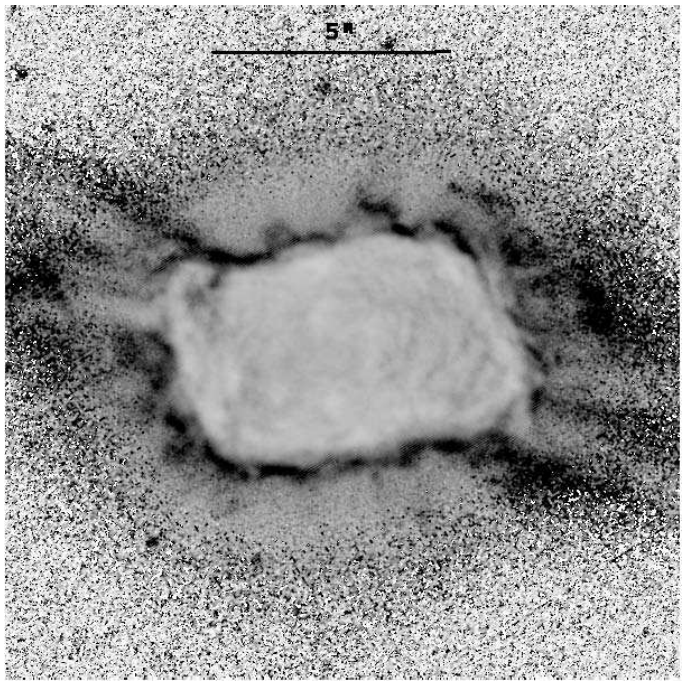} 
\caption{Apparent I([N II])/I([O III]) distribution over NGC 6741 (same WFPC2 images as Fig. 1). The sharp, inhomogeneous layer at 
low-excitation ([N II]) framing the main nebula is punched, along and close to the apparent major axis, by a series of radial [O III] rays? 
yets? penetrating into the faint, roundish halo.}  
\end{figure}

At last, the tomographic and 3-D analyses developed at the Astronomical Observatory of Padua (Sabbadin et al. 2004 and references therein) 
overcame the stumbling block of nebular de-projection, rebuilding the ionic spatial structure, and allowing us a direct comparison 
of each real, true PN with the current theoretical evolutionary models (Sch\"onberner et al. 1997, Steffen et al. 1998, Marigo et al. 2001), the 
detailed hydro-dynamical simulations (Icke et al. 1992, Frank 1994, Mellema 1997, Perinotto et al. 2004a), and the updated photo-ionization codes 
(Ferland et al. 1998, Ercolano et al. 2003).

Though the observational starting point is common - i. e. long-slit spectra -, the ``philosophy'' of tomography is just opposite of the 
conventional method. In fact, the latter compacts the spectrum along the slit (in practice, it restricts 
the nebula to a point), and gives mean, integrated results (line flux, expansion velocity, $T_{\rm e}$, 
$N_{\rm e}$, ionization etc.). Vice versa, tomography is based on a pixel-to-pixel analysis of both flux and velocity, and furnishes 
the bi-dimensional structure (in different ions) of the radial slice of nebula intercepted by the spectrograph slit. 
Later on, a 3-D rendering procedure combines all 
tomographic slices and provides the true spatial distribution of the kinematics, physical 
conditions ($T_{\rm e}$ and $N_{\rm e}$), and ionic and chemical abundances at unprecedented accuracy. 
Tomography needs spectra at high ``relative'' spatial (SS) and spectral (RR) resolutions (SS=r/$\Delta$r, r=apparent radius, 
$\Delta$r=seeing; RR=$V_{\rm exp}$/$\Delta$V, $\Delta$V=instrument spectral resolution). It is based  on the simple consideration that the position, 
depth and density of each elementary volume within an extended, 
regularly expanding nebula can be, in principle, derived from the radial velocity, FWHM and flux, respectively, of the corresponding 
emission. 

So far we have studied NGC 40 (Sabbadin et al. 2000a), NGC 1501 (Sabbadin et al. 2000b, Ragazzoni et al. 2001), NGC 6565 (Turatto 
et al. 2002), NGC 6818 (Benetti et al. 2003) and NGC 7009 (Sabbadin et al. 2004). Here we present the results for NGC 6741.

NGC 6741 (PN G033.8-02.6, Acker et al. 1992) is a compact (main body$\simeq$7''x5'', halo diameter$\simeq$15''; Curtis 1918, 
Schwarz et al. 1992), 
high-surface brightness, high-excitation (class 8, Hyung \& Aller 1997) PN with a large degree of stratification of the 
radiation. The powering star is very hot (log T$_*$$\ge$5.22, Pottasch 1981, Pottasch \& Preite-Martinez 1983, Heap et al. 1989, Tylenda et 
al. 1989, Kaler \& Jacoby 1989) and faint (m$_{\rm V}$$\simeq$19.5, Pottasch 1981, Tylenda et al. 1989, Kaler \& Jacoby 1989, Hyung \& Aller 
1997). 

The [O III] and [N II] apparent morphology of NGC 6741 (sometimes called ``Phantom Streak Nebula'') is shown in Figs. 
1 and 2 (HST-WFPC2 frames retrieved from NASA public archives).  
Note the vertical (equatorial?) inhomogeneous strip of absorbing knots in the [O III] image of Fig. 1, and in Fig. 2 the roundish halo 
and the series of weak, radial [O III] rays punching the [N II] skin along and close 
to the apparent major axis.   A multi-color HST reproduction of the nebula is given by 
Hajian \& Terzian at http://ad.usno.navy.mil/pne/gallery.html.

We immediately point out that the optical appearance of NGC 6741, the co-existence of 
ionic species with a large range of ionization potential, IP (from [O I] IP=0 eV to [Ne V] IP=97.1 eV; Aller et al. 1985, Hyung \& Aller 1997), 
and the characteristics of the central star (hot and faint) are suggestive of a recombining PN, i. e. the star has exhausted 
the hydrogen-shell nuclear burning and is rapidly fading in luminosity (and temperature); the UV flux being unable to fully ionize the 
gas, the outer nebular regions recombine, producing the faint halo (Tylenda 1986, Phillips 2000). Under many respects NGC 6741 
is very much like NGC 6565 (Turatto et al. 2002). 

To deepen the kinematics, physical conditions, ionic and spatial structure, distance and evolutionary status 
of NGC 6741, we have secured long-slit  ESO NTT+EMMI echellograms (spectral range
$\lambda\lambda$3900-7900 $\rm\AA\/$, spectral resolution $\lambda$/$\Delta$$\lambda$=R=60\,000) 
at nine position angles (PA). The spectra were reduced and analysed using our  
reconstruction 3-D technique (Turatto et al. 2002,  Benetti et al. 2003).

The results are presented in this paper, 
whose plan is as follows: Sect. 2 illustrates the 
observational procedure and the reduction method, Sect. 3 defines the kinematics of the ionized gas, Sect. 4 quantifies the 
interstellar and circum-nebular absorption, Sect. 5 provides the nebular distance, size and age, in Sect. 6 we discuss the parameters of the 
central star (temperature, luminosity, mass and evolutionary phase) 
and in Sect. 7 the nebular parameters ($T_{\rm e}$, $N_{\rm e}$, ionic mass and structure, chemical abundances, photo-ionization model), 
Sect. 8 re-builds the 3-D spatio--kinematical structure, Sect. 9 contains the general discussion, and Sect. 10 draws some conclusions.

\section{Observations and reductions}
NGC 6741 was observed with ESO NTT + EMMI (echelle mode; grating $\#$14, grism $\#3$) at nine equally spaced PA, under photometric 
sky conditions and seeing ranging between 0.50\arcsec\,and 0.70\arcsec. The spectrograph slit (1.0\arcsec \,wide 
and 30\arcsec\,long) was  
centered on the nebular image, the exciting star being invisible in the slit-viewer. 
The echellograms (exposure time 600s) cover the spectral range  $\lambda$$\lambda$3900--7900 $\rm\AA\/$ with resolution 
R$\simeq$60\,000, and provide 
the kinematical structure of the main ionic species  within the nebular slices covered by the slit. 
Bias, zero--order flat field and distortion corrections, and wavelength and flux calibrations were performed according 
to the straightforward procedure fully described by  Turatto et al. (2002). 
\begin{figure*}
   \centering \includegraphics[width=17cm]{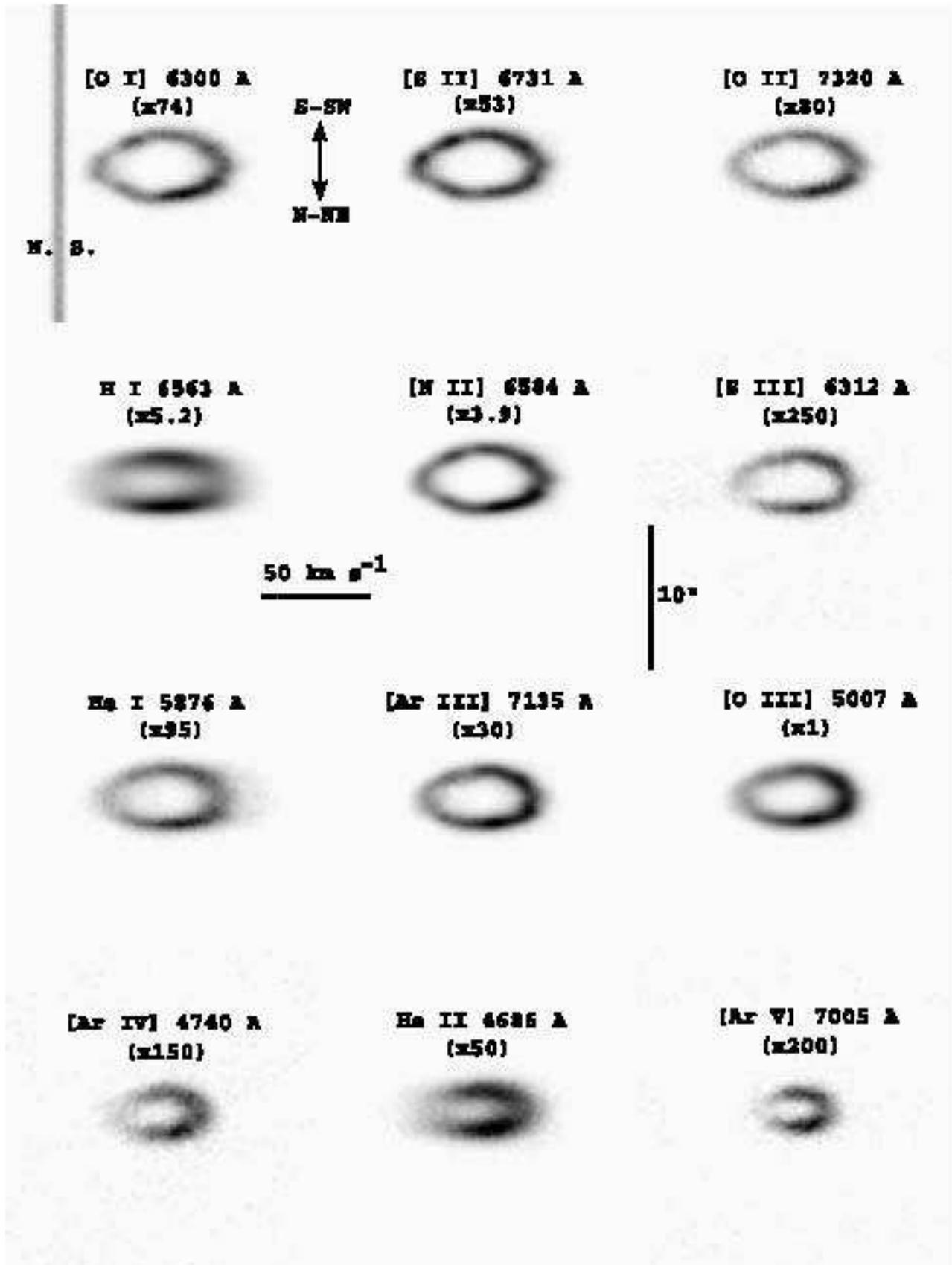}
   \caption{Detailed spectral image of twelve ionic species in NGC 6741 at PA=15$\degr$ (close to the apparent minor axis), 
arranged in order of increasing IP (top-left to bottom-right).  
The original fluxes are multiplied by the factor given in parenthesis, to make each emission 
comparable with $\lambda$5007 $\rm\AA\/$ of [O III]. The blue-shifted gas is to the left. The top-left frame also shows the [O I] 
night-sky emission at $\lambda$6300.304 $\rm\AA\/$. In the top-right frame the [O II] line at $\lambda$7319 $\rm\AA\/$, partially in 
blend with $\lambda$7320 $\rm\AA\/$, has been suppressed (for details, see the text and Fig. 5).}  
\end{figure*}
\begin{figure*}
   \centering \includegraphics[width=17cm]{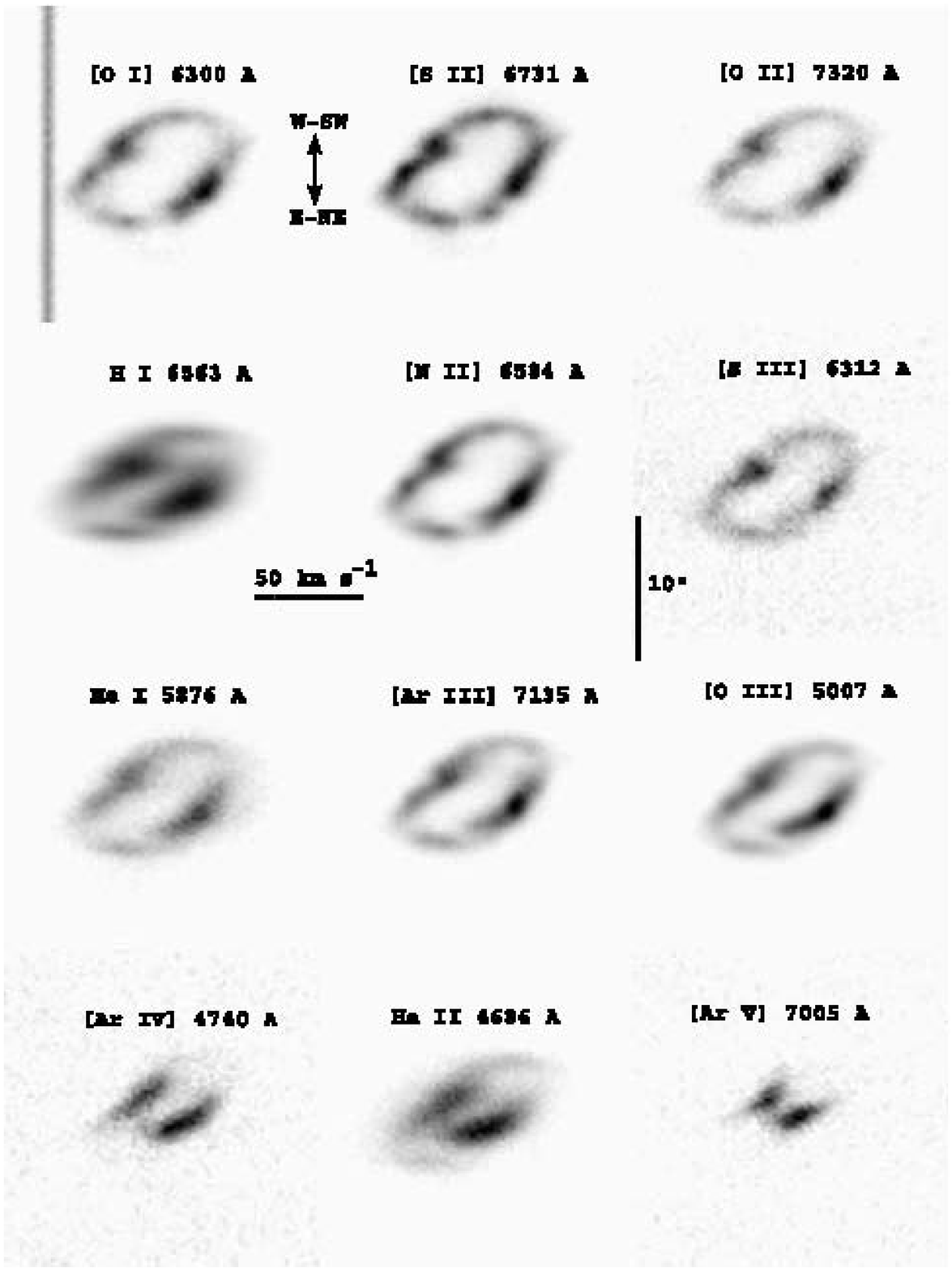}
   \caption{Same as Fig. 3, but for PA=75$\degr$ (close to the apparent major axis of NGC 6741).}  
\end{figure*}

We stress the fundamental difference between our frames, covering 80 echelle orders, and the observing procedure usually adopted for 
extended objects, inserting an interference filter to isolate a single order. The same is valid for the reduction and investigation 
methods: so far long-slit echellograms have been used to obtain, either the kinematics in a few ions (in general, [O III], H I and 
[N II]), or the ``average'' nebular properties (physical conditions, ionic and chemical abundances) integrated over the whole slit 
length. On the contrary, we proceed to a detailed, pixel-to-pixel, flux and velocity determination for a number of nebular lines, thus inferring 
the spatial distribution of the kinematics, diagnostics, ionic and total abundances at the same time. 

The richness of physical information contained in the echellograms is illustrated in Figs. 3 and 4, presenting the spectral structure 
of twelve ionic species (from [O I], IP=0 eV to [Ar V], IP=59.8 eV) at PA=15$\degr$ and 75$\degr$ (close to the apparent minor and major axes 
of NGC 6741, respectively).

These figures evidence the spectral characteristics common at all PA, in particular: (a) large stratification of 
both the radiation and kinematics (compare, e. g., the ionic sequences [O I]-[O II]-[O III] and [Ar III]-[Ar IV]-[Ar V]), and (b) 
blurred appearance of recombination 
lines (of H I, He I and He II), due to a combination of thermal motions, fine-structure and expansion velocity gradient. At the same time, 
they highlight the kinematical and physical differences at the two PA:
 \begin{description} 
\item[-] un-tilted, quite uniform emissions close to the apparent minor axis (Fig. 3), though the blue-shifted gas is systematically fainter 
than the red-shifted gas,
\item[-] tilted, inhomogeneous lines close to the apparent major axis (Fig. 4), suggestive of a dense equatorial torus and two extended and 
faint polar lobes.
\end{description}
In the original NTT+EMMI frames, $\lambda$7320.121 $\rm\AA\/$ of [O II] (top-right panel in Figs. 3 and 4; hereafter $\lambda$7320 
$\rm\AA\/$) is partially in blend with $\lambda$7319.044 $\rm\AA\/$, the bluest component of the [O II] red quartet (hereafter $\lambda$7319 
$\rm\AA\/$). Since the intensity ratio I($\lambda$7320 $\rm\AA\/$)/I($\lambda$7319 $\rm\AA\/$)=constant=3.071 in 
the $N_{\rm e}$ range here considered (De Robertis et al. 1985, Keenan et al. 1999, Sharpee et al. 
2004), the $\lambda$7319 $\rm\AA\/$ suppression was obtained in the simple fashion illustrated in Fig. 5. 

\begin{figure}
   \centering \includegraphics[width=9cm]{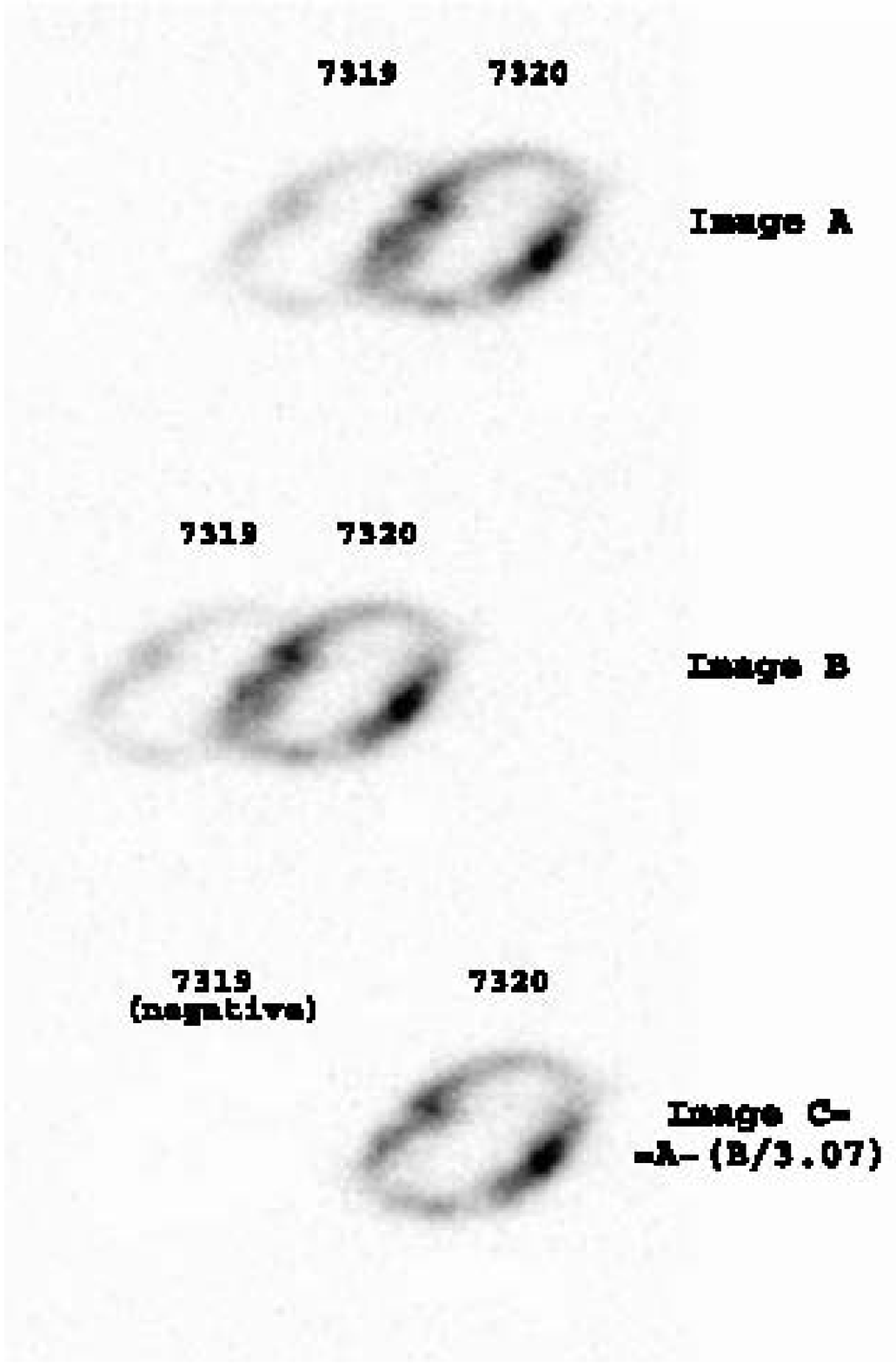}
   \caption{De-blending procedure for the [O II] $\lambda\lambda$7319-7320 $\rm\AA\/$ doublet (using IRAF packages). Top panel: Image A= part
of the echellogram (at PA=75$\degr$) centered on $\lambda$7319 $\rm\AA\/$. Middle panel: Image B= part of the echellogram (at PA=75$\degr$) 
centered on $\lambda$7320 $\rm\AA\/$. Bottom panel: Image C = Image A - (Image B/3.071), where 3.071 corresponds to I(7320)/I(7319).}  
\end{figure}
\begin{figure}
   \centering \includegraphics[width=9cm]{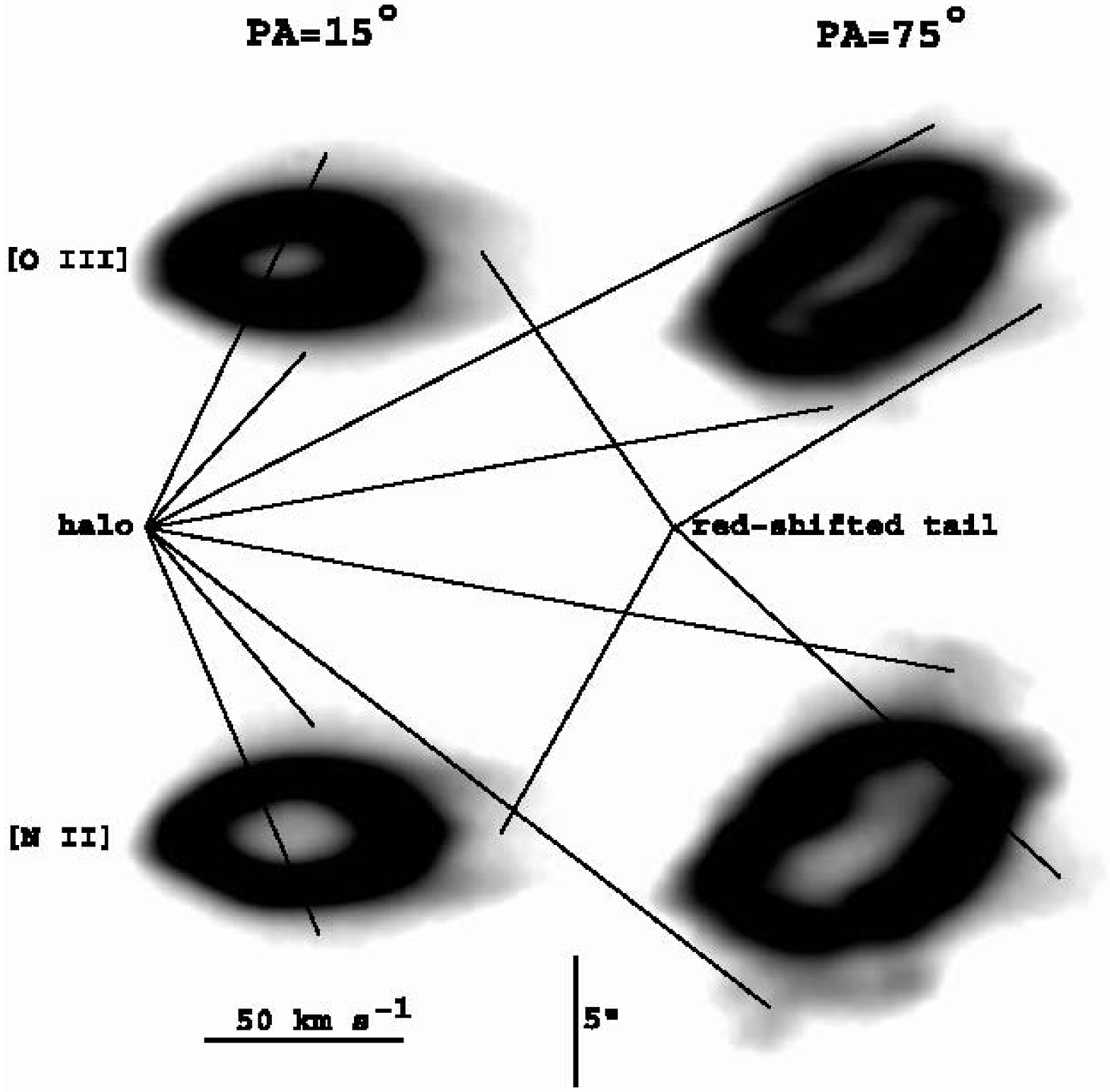}
   \caption{High-contrast [O III] and [N II] line profiles (in logarithmic scale) of NGC 6741 close to the apparent minor and major 
axes (PA=15$\degr$ and 75$\degr$, respectively) showing the faint halo and red-shifted tail. Same orientation as Figs. 3 and 4.}  
\end{figure}
We note in passing that the ``relative'' spectral and spatial resolutions of our echellograms are similar to each other:   
SS=$V_{\rm exp}$/$\Delta$V$\simeq$3 to 5 $\simeq$ RR=r/$\Delta$r. This means that the spatial information of NGC 6741 is as accurate as 
the kinematical information,  
and both are close to the lower limit for the tomographic analysis (Ragazzoni et al. 2001). Ergo, the spectral images in Figs. 3 and 4 are 
affected by non-negligible blurring agents. 

Seeing and guiding are the broadening components 
along  the spatial (i. e. vertical) axis; in our case the full-width at half maximum,  W(spat)$_{\rm seeing+guiding}$, 
results to be 0.60$\arcsec$ for PA=15$\degr$ (Fig. 3), and 0.65$\arcsec$ for PA=75$\degr$ (Fig. 4). The true, ``intrinsic'' profile of the 
nebula has:
\begin{equation}
{\rm W(spat)_{\rm intrinsic}^2= W(spat)_{\rm obs}^2 - W(spat)_{seeing+guiding}^2}. 
\end{equation}
Concerning the velocity (i. e. horizontal) axis, we must take into account:
\begin{description}
\item{(1)} instrumental resolution, corresponding to a gaussian profile with W(vel)$_{\rm EMMI}$=5.09 km s$^{-1}$ (measured in the [O I] 
night-sky line at  $\lambda$6300 $\rm\AA\/$); 
\item{(2)} thermal motions, generating a gaussian distribution with W(vel)$_{\rm thermal}$ $\simeq$21.6$\times$10$^{-2}\times$$T_{\rm e}$$^{0.5}
\times$${\rm m}$$^{-0.5}$ km s$^{-1}$, where m is the atomic weight of the element (Clegg et al. 1999 and references therein); we use the 
radial $T_{\rm e}$ profile given in Sect. 7;
\item{(3)} turbulence, i.e. random, small-scale motions; this is a very uncertain parameter for PNe, whose quantification is deferred to 
a dedicated paper (in preparation), based on very-high resolution (R$\simeq$115\,000) spectra secured with Telescopio Nazionale Galileo (TNG) 
+ SARG on a representative sample of targets. In the present case of NGC 6741, at the moment we can only infer that  W(vel)$_{\rm turb}$ is below 
10.0 km s$^{-1}$, as indicated by the 
sharpness of the spectral images of forbidden lines along the velocity (i. e. x) axis (see Figs. 3 and 4; larger turbulences should 
produce blurred spectral images of forbidden lines, similar to the H$\alpha$ one). According to the ``a posteriori'' analysis presented at the 
end of Sect. 3, in the following we will assume W(vel)$_{\rm turb}$=
3.5 km s$^{-1}$, 
in partial agreement with the general results by Neiner et al. (2000), and Gesicki et al. (2003) (but see the caveat in Sect. 3); 
\item{(4)} fine-structure (only for recombination lines); following Clegg et al. (1999),   
W(vel)$_{\rm fine-s.}\simeq$7.5 km s$^{-1}$ for H$\alpha$; moreover, we adopt W(vel)$_{\rm fine-s.}$=5.0 km s$^{-1}$ for 
$\lambda$4686 $\rm\AA\/$ (after suppression of the 20 km s$^{-1}$ blue-shifted tail with the method illustrated in Fig. 5) and  
$\lambda$6560 $\rm\AA\/$ of 
He II,  and $\lambda$5876 $\rm\AA\/$ of He I.
\end{description}
Note that thermal motions and/or fine-structure represent the main broadening factors for recombination lines, whereas 
instrumental resolution, thermal motions and turbulence are (more or less) equivalent for forbidden lines. 

The final broadening along the velocity axis is: 
\begin{equation}
{\rm W(vel)_{blur}^2= \sum_{i=1}^4 W(vel)_{i}^2},
\end{equation}
and the true, intrinsic full-width at half maximum:
\begin{equation}
{\rm W(vel)_{intrinsic}^2= W(vel)_{obs}^2 - W(vel)_{blur}^2}.
\end{equation}
W(vel)$_{\rm intrinsic}$ (in km s$^{-1}$) corresponds to the expansion velocity range of the emitting layer; it can be transformed into 
arcsec by means of the general expansion law of the ionized gas: in NGC 6741 being $V_{\rm exp}$(km s$^{-1}$)=13($\pm$1)$\times$R$\arcsec$   
(see Sect. 3), we have W(vel)$_{\rm intrinsic}$ (arcsec)= [W(vel)$_{\rm intrinsic}$ (km s$^{-1}$)]/13.

Let us consider, for example, the spectral images at PA=15$\degr$ (Fig. 3). The intrinsic FWHM in both the zvpc and cspl (as defined at the 
opening of 
Sect. 3) results to be 0.52$\arcsec$ to 0.72$\arcsec$  (forbidden lines), 0.77$\arcsec$ to 0.95$\arcsec$ (He I), 1.00$\arcsec$ to 1.15$\arcsec$ 
(He II), and 1.35$\arcsec$ to 
1.45$\arcsec$ (H I), thus confirming that: 

(a) each emitting region is extremely sharp,

(b) the whole spectral images must be carefully deconvolved: to this end we use the Richardson-Lucy algorithm (Richardson 1972, Lucy 1974), 
and a point-spread 
function given by a bi-dimensional gaussian profile characterized by W(spat)$_{\rm seeing+guiding}$ and W(vel)$_{\rm blur}$. 

So far we have considered the bright main shell of NGC 6741. At lower-flux cuts the faint halo appears (Fig. 6), whose spectral signature supports the 
recombination hypothesis (note, in particular, the broad, un-tilted halo-emission at PA=15$\degr$, and the broad, tilted 
halo-emission mimicing the kinematics of the main nebula at PA=75$\degr$).  In fact, according to the current 
evolutionary models (e. g. Sch\"onberner et al. 1997, Steffen et al. 1998, Marigo et al. 2001, Perinotto et al. 2004a), a PN halo can be:
\begin{description}
\item[-] an {\bf AGB halo}, i. e. the low-density envelope of an optically thin PN, directly ionized by the UV flux of the bright central 
star. It represents the AGB mass-loss in the pre-superwind phase, and expands at the original ejection  velocity ($V_{\rm exp} 
\simeq$10-15 km s$^{-1}$), independent on the kinematical field of the main nebula;
\item[-] a {\bf recombination halo}, corresponding to the outer, almost neutral parts of the main nebula, no more reached 
by the UV flux of a hot central star which has exhausted the H-shell nuclear burning and passed the turnover point in stellar evolution, 
rapidly fading in luminosity and temperature. The recombining layers essentially retain the general kinematical properties of 
the main thick nebula (as observed in the halo of NGC 6741).
\end{description}

Fig. 6 also shows the faint red-shifted tail present in the strongest emissions of NGC 6741 at all PA, whose nature (halo? 
interaction with the ambient ISM? early-AGB wind? instrumental scattered light?) remains unclear.

This preliminary, qualitative description enhances the complexity of NGC 6741 (a common characteristic of all PNe analysed at 
adequate spatial and spectral resolutions). Before starting a thorough investigation of the kinematical and physical 
information contained in the spectra, we underline that the general characteristics of NGC 6741 - in particular: (a) nebular 
compactness, and (b) star weakness (the stellar continuum, generally used as a position marker, is absent in the echellograms) - do make 
this nebula the ideal target to test the reliability limits of our 3-D reconstruction procedure.    

\section{The gas spatio-kinematics}
According to Sabbadin et al. (2004, and references therein), the overall spatio-kinematical properties of a regularly expanding 
nebula can be derived by combining the kinematics of the gas projected at the apparent position of the star (the ``central star 
pixel line'', cspl, of the echellograms, common at all PA) with the spatial profile at the systemic radial velocity (the ``zero 
velocity pixel column'', zvpc, at each PA).

Though the stellar continuum  is absent in the echellograms of NGC 6741, 
we obtain a satisfactory cspl-location and zvpc-alignement (within $\pm$0.1$\arcsec$) of the echelle orders 
thanks to: (a) the central, symmetrical position of the star in the broad-band HST-WFPC2 nebular images, (b) the cspl- and zvpc-calibration 
given by the continuum spectrum of the central star of NGC 7009 (observed at the same night, and with the same instrumental 
configuration). 
\begin{centering}
\begin{table*}
\caption{Peak separation in the cspl of NGC~6741}
\begin{tabular}{ccccc}
\hline
\\
Ion &IP range (eV)&&2$V_{\rm exp}$ (km/s)&\\
\cline {3-5}
\\
&& Wilson (1950)& Robinson et al. (1982) & this paper \\
\\
\hline
\\
$[$O I$]$   &  0.0-13.6  & -   &    -    & 55.1   \\
$[$S II$]$  & 10.4-23.3  & -    &  -   &   54.0\\
$[$O II$]$  & 13.6-35.1  & 44.2 & -   & 53.7  \\
H I         & $\ge$13.6  &   &   -   &44.0  \\
$[$N II$]$  & 14.5-29.6  & 42.1  & - &53.5  \\
$[$S III$]$ & 23.4-34.8  & -     & -   &48.2  \\
He I        & 24.6-54.4  & -  & -   & 48.0 \\
$[$Ar III$]$& 27.6-40.7  & -  & -   & 47.6   \\
$[$O III$]$ & 35.1-54.9  &41.6  &   42 &46.1  \\
$[$Ar IV$]$ & 40.7-59.8  & -  & -   & 33.7 \\
$[$Ne III$]$& 41.0-63.4  & 41.0 & -   & 48.2 \\
N III       & 47.4-77.5  & -  & -   & 40.0 \\
He II       & $\ge$54.4  & -  & -   & 34.0 \\
$[$Ar V$]$  & 59.8-75.0  & -  & -   & 25.0 \\
$[$Ne V$]$  & 97.1-126.2  & 0.0:   &    -  &-  \\
\\
\hline
\end{tabular}
\end{table*}
\end{centering}

The peak separations in the cspl of NGC 6741, 2$V_{\rm exp}$, are contained in Table 1 (last column), where 
ions are put in order of increasing IP.
Typical errors are 1.0 km s$^{-1}$ for the strongest forbidden emissions (like
$\lambda$4959-5007 $\rm\AA\/$ of [O III] and $\lambda$6548-6584 $\rm\AA\/$ of [N II])
to 2.0 km s$^{-1}$ for the faintest ones (in particular, 
$\lambda$6312 $\rm\AA\/$  of [S III], $\lambda$4711-4740 $\rm\AA\/$ of [Ar IV]  and $\lambda$6435-7005 $\rm\AA\/$ of [Ar V]). 
The uncertainties for recombination lines
are: 2.0 km s$^{-1}$ for $\lambda$4861-6563 $\rm\AA\/$ of H I, $\lambda$4686-6560
$\rm\AA\/$ of He II and $\lambda$4640 $\rm\AA\/$ of N III, and 1.5 km
s$^{-1}$ for $\lambda$5876-6678 $\rm\AA\/$ of He I. 

The agreement with the kinematical results given in the literature (also contained in Table 1) is quite poor, probably due to the 
compactness of NGC 6741, combined with image rotation on the slit (long-exposure Coud\'e spectra by Wilson 1950) or inadequate spatial 
resolution (circular aperture 18$\arcsec$ in diameter by Robinson et al. 1982).

\begin{figure*} \centering
\includegraphics[width=17.9cm]{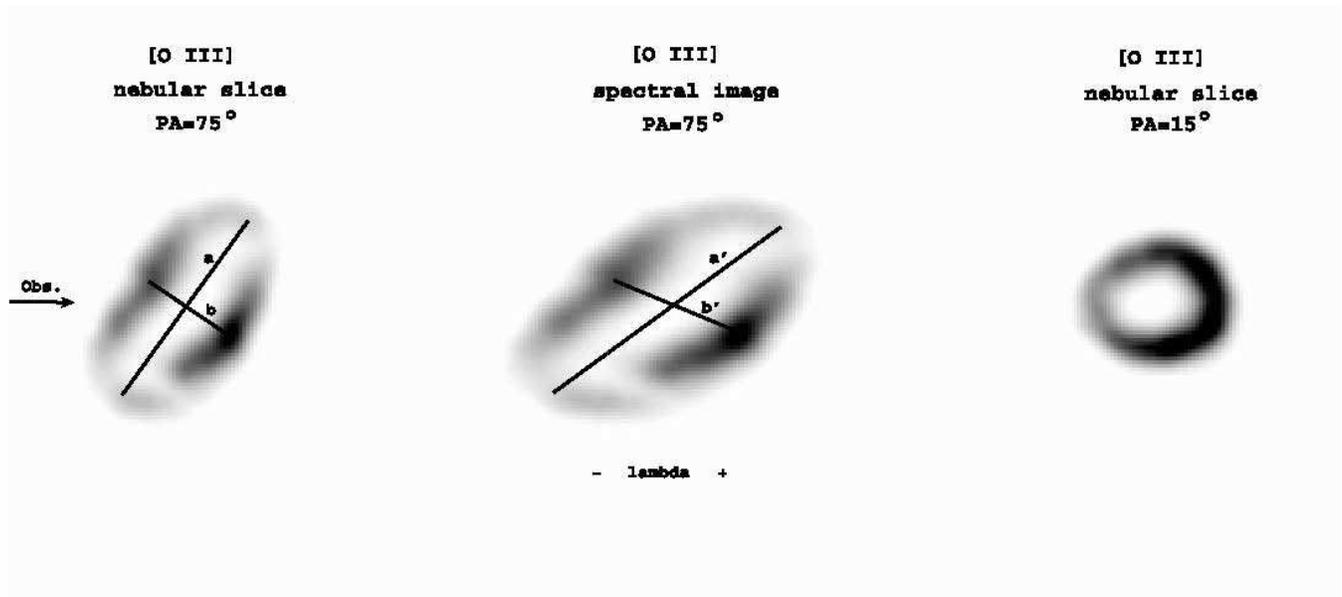} 
\caption{Nebular tomography vs. spectral image  connection. In this example we use the [O III] spectral images of NGC 6741 at PA=75$\degr$ and 
PA=15$\degr$ (same orientation as Figs. 3 and 4). 
Left panel: tomographic map of the nebular slice intercepted by a spectrograph slit aligned with the apparent major axis of an ellipsoidal PN 
denser at the equator than at the poles; a and b are the polar and equatorial axes, respectively, of the elliptical nebular slice. 
Central panel: spectral image of the nebular slice shown in the left panel; a' and b' are the polar and equatorial axes, respectively, of the 
spectral image. Right panel: [O III] tomographic reconstruction of NGC 6741 at PA=15$\degr$. See the text for details.}  
\end{figure*}

Concerning the zvpc at the nine PA of NGC 6741, the intensity peak separations, 2r$_{\rm zvpc}$, 
in the different ionic species are listed in Table 2. 

\begin{centering}
\begin{table*}
\caption{Peak separation in the zvpc at the nine PA of NGC~6741}
\begin{tabular}{ccccccccccccc}
\hline
\\
Ion&&&&2r$_{\rm zvpc}$&(arcsec)&&&\\ 
\cline{2-10}
\\
&PA=15$\degr$ &35$\degr$ &55$\degr$ &75$\degr$ &95$\degr$ &115$\degr$ 
&135$\degr$ &155$\degr$&175$\degr$\\
\\
\hline 
\\  
$[$O I$]$     & 4.3   & 4.6  & 5.7  & 7.3  & 6.8 & 7.1 &6.1 & 5.6 & 4.6 \\
$[$S II$]$    & 4.1   & 4.5  & 5.5  & 7.1  & 6.6 & 6.9 &6.0 & 5.5 & 4.4 \\
$[$O II$]$    & 4.0   & 4.3  & 5.2  & 6.6 & 6.5  & 6.6  &5.8& 5.2 & 4.3\\
HI            & 3.5   & 3.5  & 4.1  & 6.0  & 6.0 & 6.0 &5.2 & 4.5 & 3.8\\
$[$N II$]$    & 4.1   & 4.4  & 5.4  & 6.8  & 6.5 & 6.5 &5.9 & 5.2 & 4.4  \\
$[$S III$]$   & 3.6   & 4.1  & 4.6  & 6.0  & 5.7 & 6.0 &5.5 & 4.6 & 3.9 \\
He I          & 3.8   & 4.0  & 4.5  & 6.2  & 5.7 & 6.0 & 5.5& 4.6 & 4.0  \\
$[$Ar III$]$  & 3.7   & 3.8  & 4.5  & 5.9  & 5.0 & 6.0 & 5.5& 4.6&  4.1 \\
$[$O III$]$   & 3.6   & 3.7  & 4.4  & 5.3  & 4.7 & 5.5 & 5.0& 4.4 & 3.8  \\
$[$Ar IV$]$   & 2.7   & 3.2  & 3.6  & 3.9  & 3.9:  & 5.3: & 4.5:  & 3.4 &3.1  \\
$[$Ne III$]$  & 3.6   & 3.5:  & 4.0  & 4.5  & 5.0: & 5.6: & 5.0& 4.0 &3.6 \\
N III         & 3.0:   & 3.1:  & 3.6: & 4.0:  & 4.2: & 5.1: & 4.4: & 3.4:&3.2:   \\
He II         & 2.6   & 2.9  & 3.4  & 3.8  & 3.9:& 5.0:    & 3.9   & 3.2  &2.9  \\ 
$[$Ar V$]$    & 2.3   & 2.6     & 2.8     & 3.0     & 3.5:   & 4.0:   & 3.6   &  3.0  &  2.7\\
\\
\hline
\end{tabular}
\end{table*}
\end{centering}

To assemble the kinematical results of Table 1 and the spatial results of Table 2 
we can follow two ways: 
\begin{description}
\item[(I) ] ${\bf Search\,for\, the\, PA\, with\, R_{\rm zvpc} \simeq R_{\rm cspl}}$. Let us assume for the main shell of NGC 6741 the most general spatial 
structure, i. e. a tri-axial ellipsoid with axes a, b and c. 
At PA=75$\degr$ to 115$\degr$ (close to the apparent major axis, Fig. 4) the line-tilt indicates that the observer is not aligned with the major 
axis, and the overall emission structure (suggestive of a dense equatorial torus + two faint polar lobes) excludes the oblate 
(a=b) ellipsoid hypothesis. Moreover, the absence of line-tilt at PA=15$\degr$ (perpendicular to the apparent major axis, Fig. 3) means that 
either the minor axis of the ellipsoid is close to PA=15$\degr$, or the intermediate 
axis of the ellipsoid is close to PA=15$\degr$, or the ellipsoid is prolate (b=c). In all cases we conclude that, in first approximation, 
R(minor axis)$\simeq$R(intermediate axis)$\simeq$R$_{\rm zvpc}$(PA=15$\degr$)$\simeq$ R$_{\rm cspl}$, providing (through 
Tables 1 and 2): 
$V_{\rm exp}$(km s$^{-1}$)=13($\pm$2)$\times$R$\arcsec$. 
 
\item[(II)] ${\bf Spectral\, image\, deformation\, along\, the\, major\, axis}$.
Let us consider a generic, expanding ($V_{\rm exp}\propto$R) tri-axial ellipsoid denser in the equatorial region than at the poles, 
seen 
at an intermediate direction. A spectrograph slit aligned with the apparent major axis intercepts the radial slice of nebula shown in 
Fig. 7 (left panel): an ellipse with polar axis a and equatorial axis b, perpendicular to each other. The corresponding spectral 
image is an ellipse too (Fig. 7, central panel), but deformed (with respect to the tomographic map) by the telescope + 
spectrograph characteristics. Note that:

-  the polar axis (a') and the equatorial axis (b') of the spectral image are no more perpendicular to each other,

-  the original tomographic map (Fig. 7, left panel) can be obtained by means of a simple compression of the spectral 
image (Fig. 7, central panel) along the x (i. e. velocity) axis, till having a'$\bot$b'. $\footnote {This visual effect can be seen by 
rotating Fig. 7 around the y axis.}$

In practice, and  reversing the foregoing procedure, we compress the [O III] spectral image of NGC 6741 at PA=75$\degr$ 
(Fig. 7, central panel) by a factor 1.60 along the x axis, thus correcting for the spectral deformation introduced by NTT+EMMI 
on the nebular slice intercepted by the slit, and obtaining:

- the tomographic map shown in Fig. 7 (left panel),

- the expansion law  $V_{\rm exp}$(km s$^{-1}$)=13($\pm$1)$\times$R$\arcsec$. 

This is the same law given by method (I). In fact, 
the compression factor along the x axis being the same at all PA, we can repeat the procedure for the [O III] spectral image of NGC 6741 
at PA=15$\degr$ (close to the apparent minor axis; Fig. 3), thus recovering the tomographic map shown in the right panel of Fig. 7, i. e. an almost 
circular ring with  R$_{\rm zvpc}$(PA=15$\degr$)$\simeq$ 
R$_{\rm cspl}$, QED! $\footnote {quod erat demostrandum; Latin for: which was to be proved.}$
\end{description}
The detailed cspl--zvpc relation  for NGC 6741 at PA=15$\degr$, shown in Fig. 8, indicates that:

a) on the whole, the ionized gas follows Wilson's law: the high-excitation 
zones expand more slowly than the low-excitation ones, and a positive correlation exists between the expansion velocity and 
the size of the monochromatic image (Wilson 1950); 

b)  the range of both 2r$_{\rm zvpc}$ and 2$V_{\rm exp}$ is quite large (2.3 to 4.3 arcsec 
and 25.0 to 55.1 km s$^{-1}$, respectively), suggesting  a broad radial density profile and  large stratification of the radiation; 

c) the general expansion law, $V_{\rm exp}$(km s$^{-1}$)=13$\times$R$\arcsec$, fails in the innermost, highest ionization layers 
marked by the [Ar V] emissions: they expand slower than expected. 

Point c) is quite peculiar among PNe, and deserves some comments. Deceleration is present (and clear) in both 
the $\lambda$6435 $\rm\AA\/$ and $\lambda$7005 $\rm\AA\/$ [Ar V] lines at all PA of 
NGC 6741 (except along and close to the apparent major axis, where the open-ended structure of the [Ar V] spectral image prevents any 
conclusion; see 
Fig. 4, lower-right panel), whereas no evidence of deceleration appears in other high ionization species, like He II (due to the blurred 
emissions) and N III ($\lambda$4640 $\rm\AA\/$ is too faint). 
Though a detailed spatio-kinematical study at even higher ionization stages appears indispensable - for example in the forbidden line 
of Ne$^{+4}$ (IP=97.1 eV) at $\lambda$3425 $\rm\AA\/$ (which is outside our spectral range) - a support to the 
deceleration hypothesis comes from the classical work by Wilson (1950), who obtained $V_{\rm exp}$[Ne V]$\simeq$0 km s$^{-1}$ 
(quite uncertain). 

We believe NGC 6741 is the first PN showing clear evidences of deceleration in the highest ionization layers (the second candidate  
being IC 418: a preliminary analysis of our NTT+EMMI echellograms, taken at six PA, suggests the probable presence of infalling gas).

Vice versa, recent results by Gesicki \& Zijlstra (2003, 3 PNe) and Gesicki et al. (2003, 14 PNe) - based on high-resolution emission 
profiles integrated 
along the slit, and spherical shell hypothesis for the emitting gas - indicate  
that acceleration is a quite common property of the PN innermost layers (i. e. ``U''-shaped expansion profile), due to the dynamical contribution 
by the shocked, hot wind 
from the central star.  Gesicki \& Zijlstra (2003) 
adopted $\lambda$4686 $\rm\AA\/$ of He II as diagnostic of the high-ionization kinematics, whereas Gesicki et al. (2003) used 
$\lambda$5007 $\rm\AA\/$ of [O III] (in a few cases $\lambda$6560 $\rm\AA\/$ of He II).

The following ${\bf caveat}$ are in order:

1) the spherical shell assumption is wide of the mark;

2) $\lambda$5007 $\rm\AA\/$ of [O III] (IP range 35.1 to 54.9 eV) is a poor diagnostic of the highest ionization strata (except for very-low 
excitation PNe);

3) the recombination lines of hydrogen and helium do suffer severe blurring effects (thermal motions, fine-structure and expansion velocity 
gradient across the nebula combined with the small number of ionization stages of H and He), introducing spurious kinematical results.

In particular, $\lambda$4686 $\rm\AA\/$, He II Paschen $\alpha$, consists of thirteen fine-structure components spread in the 
$\lambda$$\lambda$4685.377--4685.918 $\rm\AA\/$ range (corresponding to $\Delta$V=34.6 km s$^{-1}$), with a strong blue-shifted tail (see Figs. 
3 and 4). 
In the case of $\lambda$6560 $\rm\AA\/$, He II Pickering $\beta$, the fine-structure components are even nineteen, covering the  
$\lambda\lambda$6559.769--6560.209 $\rm\AA\/$ spectral range ($\Delta$V=20.1 km s$^{-1}$).

To further test this point, let us consider the emission line profiles integrated along the slit of NGC 6741 at PA=15$\degr$. 
The results for $\lambda$4686 $\rm\AA\/$ and $\lambda$6560 $\rm\AA\/$ of He II, and the ionic sequence of argon - [Ar  III] at $\lambda$7135 
$\rm\AA\/$, [Ar  IV] at $\lambda$4740 $\rm\AA\/$ and [Ar V] at $\lambda$7005 $\rm\AA\/$ - are shown in Fig. 9; they confirm that:

a) spherical symmetry is a simplistic assumption,

b) the He II recombination lines fail to fit the well-defined stratification of both the 
radiation and expansion present in the ionic sequence of argon,  
FWHM(He II, IP$>$54.4 eV) being intermediate between FWHM([Ar III], IP range 27.6-40.7 eV) and FWHM([Ar IV], IP range 40.7-59.8 eV), 
whereas, according to the detailed kinematical results presented in this section, we would expect 
FWHM(He II) $\le$ FWHM([Ar IV]) $<<$ FWHM([Ar III]).
 

The same discrepancy occurs when considering either more parameters (like full-width at 10\% maximum flux), or different ionic 
sequences, or further PA of NGC 6741, or other PNe of our sample.

A direct confirmation of the misleading kinematical results - in particular, the high-velocity of the innermost layers - provided by the 
combination (spectral profile of recombination lines integrated 
along the slit) + (spherical symmetry assumption) comes from a comparative analysis of the emission line profiles contained in 
Gesicki \& Zijlstra 
(2003) and Gesicki et al. (2003): in all cases (17 PNe) the forbidden lines of the ionic species with the highest IP, also show the 
sharpest emission profile (as expected for a simple, positive $V_{\rm exp}$ vs. radius relation).

All this questions the general validity of the spatio-kinematical studies based on the spherical symmetry assumption and/or spectral 
profiles of recombination lines integrated along the slit, and weakens (or even cancels) the reliability of their results on turbulence, 
radial kinematics, 
matter distribution and ionization, whose quantification needs detailed studies at high spatial and spectral resolutions. 

  \begin{figure} \centering
  \includegraphics[width=9.5cm, height=9.5cm]{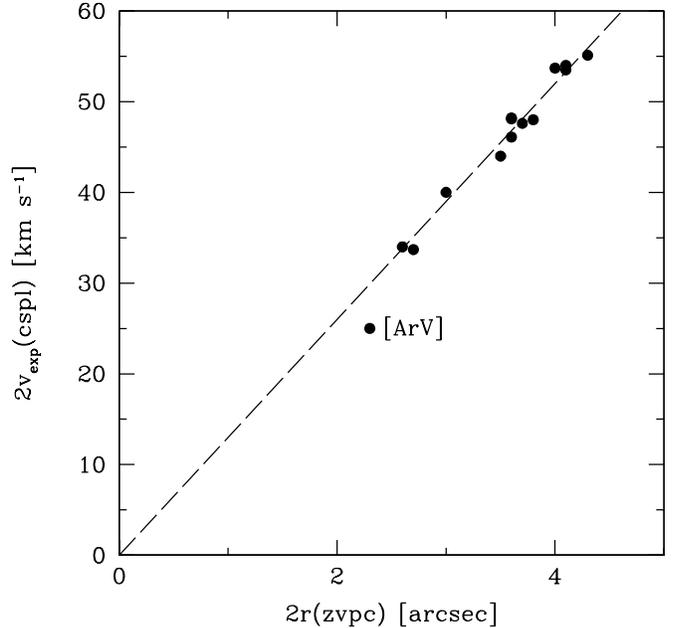}
  \caption{The cspl--zvpc relation  for NGC 6741 at PA=15$\degr$, superimposed to the adopted expansion law $V_{\rm exp}$ (km s$^{-1}$)
=13$\times$R$\arcsec$, which is valid across the whole nebula, except in the innermost, highest-ionization, decelerated regions marked by [Ar V].}  
\end{figure}

  \begin{figure} \centering
  \includegraphics[width=9.5cm, height=9.5cm]{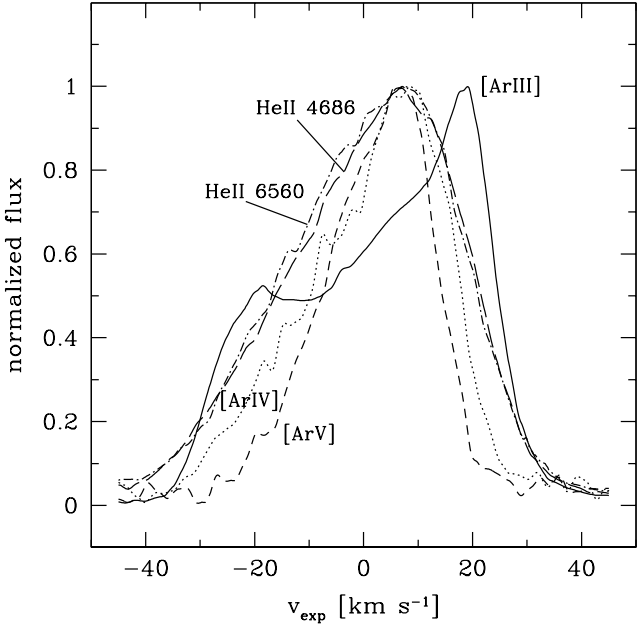}
  \caption{Selected emission line profiles integrated along the slit for NGC 6741 at PA=15$\degr$, showing the spurious kinematical 
results provided by the recombination lines of He II.  
Symbols: continuous line= [Ar  III] at $\lambda$7135 
$\rm\AA\/$; dotted line= [Ar  IV] at $\lambda$4740 $\rm\AA\/$; short-dashed line= [Ar V] at $\lambda$7005 $\rm\AA\/$; long-dashed line= 
He II at $\lambda$4686 
$\rm\AA\/$; dotted-dashed line= He II at  $\lambda$6560 $\rm\AA\/$. For details, see the text.}  
\end{figure}

Let us go on. Fig. 10 (multi-color map only in the ``free'' electronic version of the paper, for lack of funds) shows the complete velocity field at 
the nine 
observed PA of NGC 6741. We select He II, [O III] and [N II] as markers of 
the high, medium and low-excitation regions, respectively. 
These position--velocity (P--V) maps are relative to the systemic heliocentric velocity
of the nebula, $V_{\rm rad \odot}$= +39.6($\pm1.0$) km s$^{-1}$, corresponding
to $V_{\rm LSR}$= +56.4($\pm$1.0) km s$^{-1}$, and are scaled (along the x axis) according to $V_{\rm exp}$ (km s$^{-1}$)$\simeq$13$\times$R$\arcsec$, 
i.e. they reproduce the tomographic maps of the nebular slices covered by the spectrograph slit.  

Fig. 10 highlights the kinematical complexity and large stratification of the radiation within NGC 6741. The main nebula consists 
of an almost-prolate ellipsoid (a$\simeq$7.4$\arcsec$, a/b$\simeq$1.8, a/c$\simeq$2.0), whose major axis (projected at PA$\simeq$95$\degr$) 
forms an angle of 55($\pm$3)$\degr$ with the line of sight: on the whole, the eastern part of the nebula is approaching the observer, and the 
western part receding.

All this throws new light on two specific fields: (I) nature of the curious skip of absorbing knots present in Fig. 1 (upper panel), and 
(II) turbulent motions. 

{\bf (I) skip of absorbing knots}: the central location, 
orientation (vertical), and light curvature (concavity towards East) are suggestive of an inhomogeneous belt of neutral matter (gas + 
molecules + dust) embedding the dense ionized gas of the equatorial regions. The amount of circum-nebular neutral gas 
can be estimated from the [O III] flux-depletion suffered by the underlying ionized nebula, being:

\begin{equation}
\log \frac{{\rm F}(5007)_{\rm off-knot}}{{\rm F}(5007)_{\rm on-knot}}=k_{5007}\,-\,k_{\rm H_{\beta}} + 
c({\rm H}\beta)_{\rm circum-nebular},
\end{equation}
where k$_{\lambda}$ is the extinction coefficient (Seaton 1979), and c(${\rm H}\beta)_{\rm circum-nebular}$ the logarithmic extinction at 
H$\beta$ 
caused by the local absorbing matter. 

Intensity scans in the neighbourhood of the deepest knots provide $\frac{{\rm F}(5007)_{\rm off-knot}}{{\rm F}(5007)_{\rm on-knot}}$  
up to 1.4 ($\pm$0.1), i. e. 
c(${\rm H}\beta)_{\rm circum-nebular}$ up to 0.18 ($\pm$0.03). Assuming a ``normal'' gas-to-dust ratio (Spitzer 1978, Bohlin et al. 1978) and 
c(${\rm H}\beta)$=1.48$\times$E(B-V) (Acker 1978), we have:

\begin{equation}
n{\rm (H I})= \frac{4.8\times 10^{21}\times{\rm E(B-V)}_{\rm circum-nebular}}{\Delta{\rm l}},
\end{equation}

where $n{\rm (H\,I)}$ is the H I density (atoms cm$^{-3}$), and $\Delta{\rm l}$ the radial thickness (cm) of the absorbing layer. For D=2000 
pc (see Sect. 5) and $\Delta{\rm l}$=[r(halo) - r(main nebula at PA=15$\degr$)]/2$\simeq$2.0$\arcsec$ (from Figs. 1 and 2), we obtain an 
indicative value of $n{\rm (H\,I)}\simeq$7$\times$10$^3$ atoms cm$^{-3}$. 
Such a high density of the circum-nebular matter in NGC 6741 is a further support to the recombination hypothesis outlined in 
the previous sections.

{\bf  (II) turbulent motions}. In Fig. 10 the tomographic map close to the apparent minor axis (i.e. at PA=15$\degr$) being almost circular 
and quite homogeneous, we can assume {\rm W(spat)$_{\rm intrinsic}$}(zvpc at PA=15$\degr$) $\simeq$ 
{\rm W(vel)$_{\rm intrinsic}$}(cspl), thus inferring {\rm W(vel)$_{\rm turb}$} (through Eqs. (1) to (3), $V_{\rm exp}$(km s$^{-1}$)
=  13$\times$R$\arcsec$, and the $T_{\rm e}$ radial profile given in Sect. 7.1). We overlook 
the recombination lines of hydrogen and helium (dominated by thermal motions and/or fine-structure), and consider the strongest 
forbidden lines. The analysis of $\lambda$6300 $\rm\AA\/$ ([O I]), $\lambda$6731 $\rm\AA\/$ ([S II]), $\lambda$6584 $\rm\AA\/$ ([N II]), 
$\lambda$7135 $\rm\AA\/$ ([Ar III]), and $\lambda$5007 $\rm\AA\/$ ([O III]) provides {\rm W(vel)$_{\rm turb}$}= 3.5($\pm$2.0) km s$^{-1}$, 
with no evident relation to the ionization degree. Thus, in spite of the crude assumptions and wide uncertainties, we conclude 
that turbulent motions in NGC 6741 are quite modest.

\begin{figure*} \centering
\includegraphics[width=17.8cm]{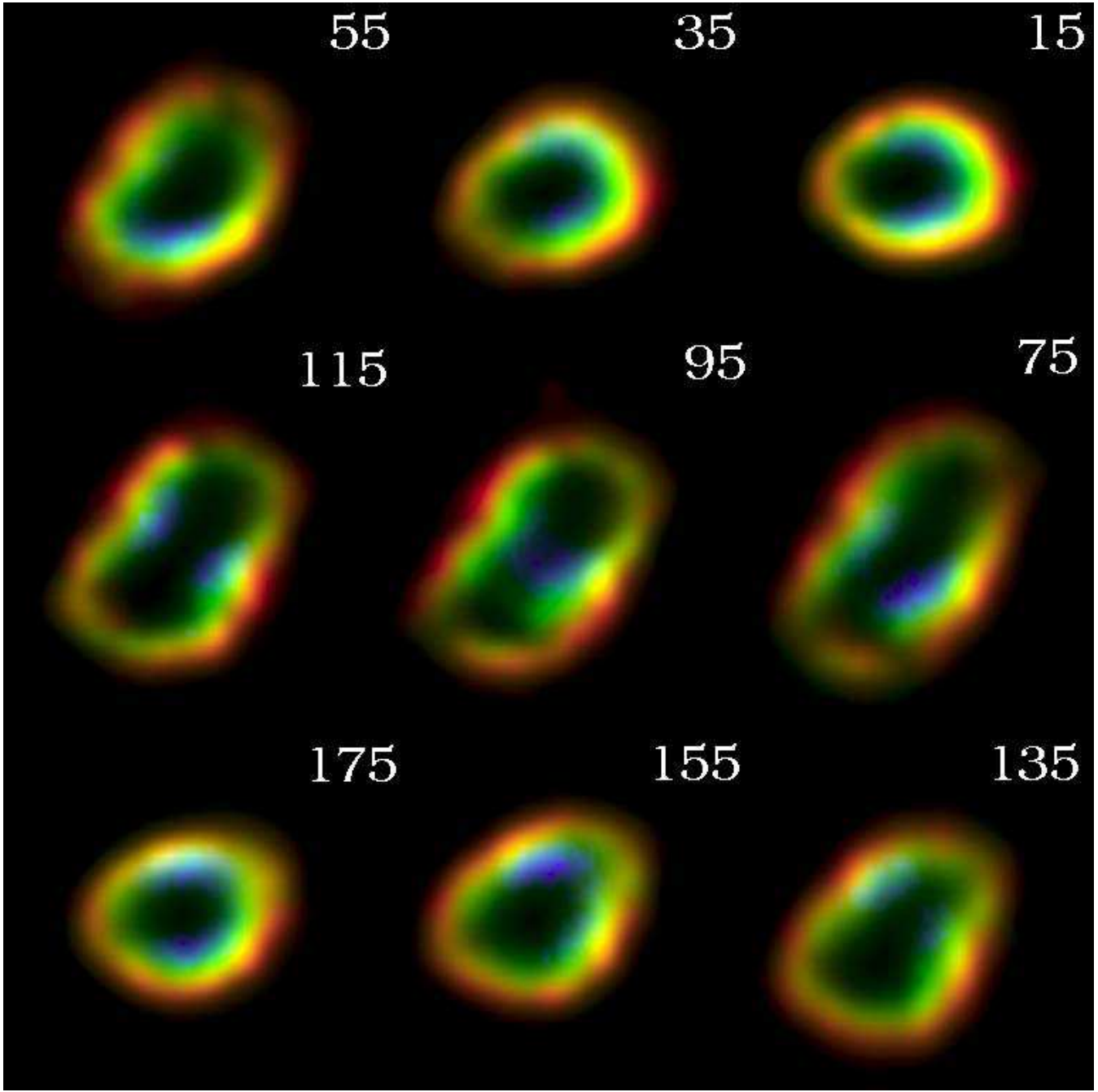} 
\caption{Combined position--velocity maps in the nine observed PA of NGC 6741 at high (He II), medium ([O III]) and low ([N II]) excitation  
(multi-color maps in the electronic version of the paper; blue=He II, green=[O III], and red=[N II]), scaled according to the relation $V_{\rm exp}$ 
(km s$^{-1}$)=13$\times$R$\arcsec$. The 
orientation of these tomographic maps is the same of Figs. 3 and 4.}  
\end{figure*}


\section{The absorption (interstellar + circum-nebular)}
In general, the observed line intensities must be corrected for absorption according to:
\begin{equation}
\log \frac{{\rm I}(\lambda)_{\rm corr}}{{\rm I}(\lambda)_{\rm obs}}=k_{\lambda}\times c({\rm H}\beta)_{\rm tot},
\end{equation}
where k$_{\lambda}$ is the extinction coefficient (Seaton 1979), and c(${\rm H}\beta)_{\rm tot}$ the logarithmic extinction at H$\beta$ 
given by:
\begin{equation}  
c({\rm H}\beta)_{\rm tot}=c({\rm H}\beta)_{\rm interstellar} + c({\rm H}\beta)_{\rm circum-nebular}
\end{equation}

with c(${\rm H}\beta)_{\rm circum-nebular}$=0 for an optically thin, fully ionized, density bounded nebula. Decidedly, this is not 
the case of NGC 6741, 
which is optically thick, ionization bounded, and wrapped up in a dense cocoon of almost-neutral matter. 

The extinction estimates reported in the literature (mean values along the nebular slice covered by the spectrograph slit) cluster 
around c(${\rm H}\beta)_{\rm tot}$=1.05 (Kaler \& Lutz 1985, Kaler \& Jacoby 1989, Cahn et al. 1992, Hyung \& Aller 1997). 

To disentangle the complex absorption over NGC 6741, we apply  the 
H$\alpha$/H$\beta$ analysis of the whole spectral image, as introduced by Turatto et al. (2002). Fig. 11 shows the F(H$\alpha$)/F(H$\beta$) 
isophotal contours superimposed to the H$\beta$ spectral image for three representative PA of NGC 6741: 
at PA=15$\degr$ (close to the apparent minor axis; untilted H$\beta$ spectral image) F(H$\alpha$)/F(H$\beta$) increases outwards along the 
spatial (i. e. y) axis, peaking beyond the top of the H$\beta$ flux. 
At PA=55$\degr$ (intermediate PA; tilted H$\beta$ spectral image) the F(H$\alpha$)/F(H$\beta$) raise along the y axis is overwhelmed by the 
broad maximum at S-W of the nebular centre (at the expected position of the absorbing belt visible in Fig. 1, upper 
panel). The same occurs at PA=75$\degr$ (close to the apparent major axis; tilted H$\beta$ spectral image), where F(H$\alpha$)/F(H$\beta$) 
peaks at W-SW of the centre, in correspondence of the equatorial absorbing belt.

\begin{figure*} \centering
\includegraphics[width=18cm,height=9cm]{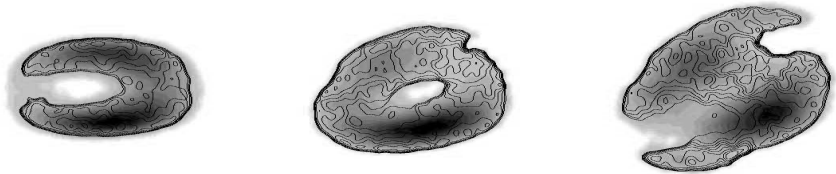} 
\caption{F(H$\alpha$)/F(H$\beta$) 
isophotal contours superimposed to the H$\beta$ spectral image for three representative PA of NGC 6741. Left panel: PA=15$\degr$ 
(close to the apparent minor axis); central panel: PA=55$\degr$ (intermediate direction); right panel: PA=75$\degr$ (close to the apparent 
major axis). Same orientation as Figs. 3 and 4. The isophotal contours cover the range 5.20 (the outermost) to 7.60, with a constant 
step of 0.30. }
\end{figure*}

Summing up the H$\alpha$/H$\beta$ intensity distribution at the nine observed PA of NGC 6741, we infer that:
\begin{description}
\item[a)] c(${\rm H}\beta)_{\rm interstellar}$=0.95 ($\pm$0.05),
\item[b)] c(${\rm H}\beta)_{\rm circum-nebular}$ is 0.10 ($\pm$0.03) in the central region of the nebular image (out of the equatorial absorbing 
belt), and raises up to 0.20 ($\pm$0.05) within the equatorial absorbing belt, and at the edge of the ionized zone.
\end{description}
Moreover, a modest decrease of F(H$\alpha$)/F(H$\beta$) in the innermost regions (out of the equatorial absorbing belt) is suggestive of 
a local increase of $T_{\rm e}$.
 
Last, an indicative value of the total neutral mass embedding the ionized nebula (through Eq. (5) and using simple geometrical considerations) 
is M$_{\rm neutral}$$\simeq$0.20 ($\pm$0.05) M$_\odot$.

Though the small angular size prevents a deeper analysis, these results do represent the umpteenth sign of the complex structure and peculiar 
evolutionary phase of NGC 6741, and will be a precious support in the determination of the topical parameter, i. e. distance.

\section{The nebular distance, size and age}
The ``statistical'' distance of NGC 6741, provided by two dozen catalogues using different methods and assumptions, is:
\begin{description}
\item[]$<$D$>$(Shklovsky)$\simeq$3300($\pm$1000) pc
\item[]$<$D$>$(ionized mass--radius relation)$\simeq$1500($\pm$700) pc 
\item[]$<$D$>$(other methods)$\simeq$2000($\pm$1000) pc, 
\end {description}
where $<$D$>$(Shklovsky) represents an upper limit of the true distance, NGC 6741 being an optically thick, ionization bounded PN.

Individual values reported in the literature are: 1300 pc (nebular radial velocity combined with the circular law of galactic rotation; 
Acker 1978), 2100 pc (Acker 1978) and 1400 pc (Pottasch 1983) (both based on the large-scale galactic interstellar absorption map by Lucke 1978), 
and 1500 pc (color-excess vs distance relation for early-type stars within 1.5$\degr$ of the nebula; Kaler \& Lutz 1985). 

We tried to determine the nebular parallax by combining the expansion velocity field with the angular expansion measured in 
first- and second-epoch HST-WFPC2 frames. The NASA 
public archives contain 21 images of NGC 6741, taken at two epochs separated by 2.97 years (programs GO 7501 and GO 8773; P. 
I. Arsen Hajian). The [O III] and [N II] multiple exposures (WFPC2 central planetary camera, PC; pixel size=0.0455 arcsec) were co-added, 
corrected for optical distortions, aligned  and rotated using 
IRAF packages (Reed et al. 1999; Palen et al. 2002).

No apparent image-shift is detected, i. e. image-shift$\le$$\frac{1}{2}$pixel, and $\frac{d\theta}{dt}$$\le$8$\times$10$^{-3}$ arcsec yr$^{-1}$. 
For an optically thin, fully ionized PN this implies D$\ge$1400 pc, being:
\begin{equation}
  {\rm D(pc)}=\frac{0.211 V_{\rm exp} {\rm (km\ s^{-1})}}{\frac{d\theta}{dt} {\rm (arcsec\ yr^{-1})}}.
\end{equation}
But NGC 6741 is optically thick (probably recombining), and  Eq. (8) cannot be applied, since in this case we should compare the expansion velocity 
of the ionized gas with the angular expansion of the ionization front. To be noticed that, in the limit case of a nearby nebula in a 
very deep recombination phase, we even expect a detectable contraction of the ionization front. Thus, we infer (a mere speculation, at the 
moment) that D(NGC 6741)$\ge$1500 pc, and/or the nebula is at the end of the recombination phase (or at the reionization start). 

As a last chance for the nebular distance determination, we decided to enrich (and sharpen, if possible) the broad color-excess vs. D 
relation given by Kaler \& Lutz (1985) with the recent bibliographic reports. 
The scanning of the NGC 6741 neighbourhood with the SIMBAD facilities of CDS (Centre de Donn\'ees astronomiques de Strasbourg) gave fruitful 
results; besides two dozen (low-weight) field stars with accurate photometry and spectral type (in most cases the luminosity class is absent; 
we assume a luminosity class V), we have identified two important distance markers:

 - the open cluster OCL 88 (at an apparent distance of 15.4' from NGC 6741), characterized by E(B-V)=1.0 and D=3000 pc (Sagar \& Griffiths 1998, 
and referenced therein),

- the classical Cepheid V336 Aql (apparent distance from NGC 6741=37.0'), with log P=0.8636 days, $<$m$_{\rm V}$$>$=9.875, E(B-V)=0.61--0.64, and 
D=1800--2100 pc (Pel 1976, Feast \& Whitelock 1997, Metzger et al. 1998, Berdnikov et al. 2000).

The improved E(B-V) vs. D relation in the direction of NGC 6741 is shown in Fig. 12: the interstellar absorption starts at D$\simeq$ 200--300 pc, 
and gradually increases up to 3000 pc (and beyond). This is in partial disagreement with the literature reports. We recall that NGC 6741 is 
within the Aquila-Scutum cloud, a galactic region characterized by a high, quite uniform star density, suggestive of a low oscuration. 
According to Forbes (1985), it shows an almost immediate climb at A$_{\rm V}$=2.3 mag. within 300 pc; beyond this distance, there is little or no 
increase of extinction out to D$\simeq$6000 pc.

\begin{figure} \centering
\includegraphics[width=9.5cm, height=9.5cm]{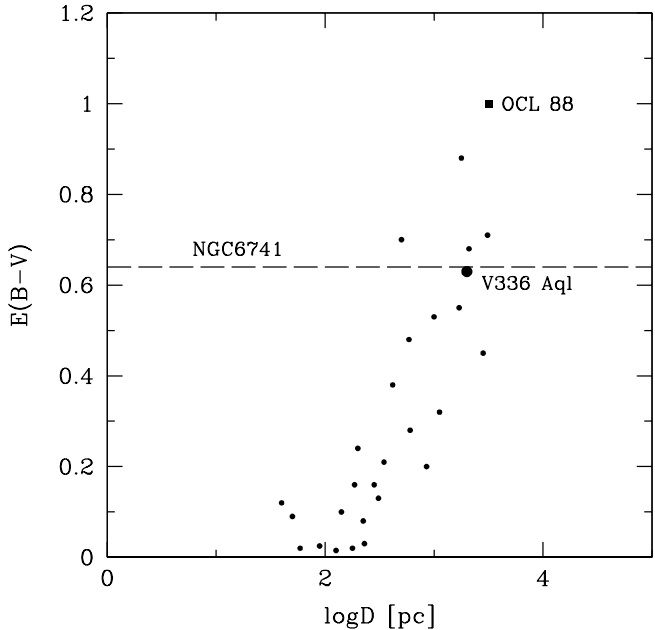} 
\caption{Color-excess vs. distance relation in the direction of NGC 6741. Symbols: small dots= field stars; filled square= open cluster OCL 88; 
large dot= classical Cepheid V336 Aql.
}
\end{figure}

To test this point, let us consider the four more PNe projected within the Aquila-Scutum cloud (at an apparent distance from 
NGC 6741 lower than 90'). They are CBSS 3, K 3-19, M 1-66 and K 3-20 (we reject a fifth object, Sp 2-151, because of the unreliable line 
intensities - in particular F(H$\alpha$)/F(H$\beta$)$\simeq$1.0 - and peculiar spectral-type A for the exciting star reported by Acker et al. 
1992, probably due to the partial blending with a relatively bright, m$_{\rm R}$$\simeq$13.0, field star). The four selected PNe in the 
Aquila-Scutum cloud do have 1.5$\le$c(${\rm H}\beta)$$\le$2.2 (line intensities by Acker et al. 1992, and Cappellaro et al. 1994, for 
$T_{\rm e}$=10$^4$ K and $N_{\rm e}$=2$\times$10$^3$ cm$^{-3}$), i. e. 3.2$\le$A$_{\rm V}$$\le$4.7, and, using the oscuration law by Forbes 
(1985), D$>$ 6000 pc for all four nebulae.

The last result appearing quite improbable, and, in our opinion, questionable, in the following we will adopt the 
color-excess vs. 
distance relation shown in Fig. 12. It provides D(NGC 6741) = 2000 ($\pm$300) pc.

The linear size of the main nebula is 0.036 pc x 0.020 pc x 0.018 pc (major, intermediate and minor semi-axes, respectively), whereas both 
the spherical recombining halo and the [O III] rays punching the [N II] skin close to the apparent major axis extend up to 0.080 pc. 

The ``kinematical'' age of NGC 6741, 
t$_{\rm kin}$=R/$V_{\rm exp}$, is about 750 years, and the ``true'' age t$_{\rm true}$$\simeq$2R/ [$V_{\rm exp}$(today)+$V_{\rm exp}$(AGB)]$\simeq$ 
1400 years. 

Summing up: NGC 6741 is a very young, compact, high surface brightness (i.e. dense) PN. The combination of: (a) high excitation 
(up to [Ne V]) of the 
innermost layers, (b) low ionization skin, and (c) almost neutral, large density halo, is indicative of a very hot central 
star at low luminosity. A deep analysis of the powering engine is in order.

\section{The central star parameters}
The HST-WFPC2 frames of NGC 6741 taken through the broad-band filter F555W (central wavelength=5407 $\rm\AA\/$, bandwidth=1236 $\rm\AA\/$) 
provide m$_{\rm V}$=20.09 ($\pm$0.05), where the un-known star color is the main source of inaccuracy. Previous ground-based estimates reported 
in the 
literature are: 19.5 (Pottasch 1981), 17.6 (Tylenda et al. 1989), 19.16 (Kaler \& Jacoby 1989) and 19.26 (Hyung \& Aller 1997) for 
m$_{\rm V}$, and $>$20.2 (Gathier \& Pottasch 1988) and 18.2 (Tylenda et al. 1989) for m$_{\rm B}$.

The H I and He II Zanstra temperatures are given by the stellar magnitude, combined with both the total H$\beta$ nebular flux, 
log F(H$\beta$)$_{\rm obs}$=-11.32 ($\pm0.03$) mW$\times$m$^{-2}$ (Kaler \& Lutz 1985, Kaler \& Jacoby 1989, Acker et al. 1991, 
Cahn et al. 1992, this paper), and the flux ratio F($\lambda$4686
${\rm \AA}$)/F(H$\beta$)=0.40($\pm0.03$) (Aller at al. 1985, Kaler \& Jacoby 1989, Hyung \& Aller 1997, this paper). 

We obtain 
log(T$_{\rm Z}$H I)=5.33($\pm0.07$) and log(T$_{\rm Z}$He II)=5.23($\pm0.07$), 
thus confirming the peculiarity already reported by Heap et al. (1989) and Hyung \& Aller (1997), i. e. the Zanstra discrepancy is reversed. 
T$_{\rm Z}$H I$>$T$_{\rm Z}$He II is a typical signature of recombining PNe, e. g.  NGC 6565 (Turatto et al. 2002), where the 
ionization and thermal structure are out of equilibrium, and the recombination processes dominate. These are faster for the 
higher-ionization species (Tylenda 1986, Stasinska 1989, Marten \& Szczerba 1997); thus, in the following, we will adopt 
T$_*$$\simeq$T$_{\rm Z}$He II$<$T$_{\rm Z}$H I.
 
The stellar luminosity (using D=2000 pc, and the bolometric corrections by Sch\"onberner 1981) results to be log L$_*$/L$_\odot$=2.75($\pm0.15$). 
 
The high temperature and low luminosity of the star, added to the short nebular age, suggest that the stellar mass, 
M$_*$, is larger 
than the average value ($\simeq$0.60 M$_\odot$) of the PNe nuclei. According to the evolutionary tracks by 
Sch\"onberner (1981, 1983), Iben (1984),  
Wood \& Faulkner (1986), Bl\"ocker \& Sch\"onberner (1990), Vassiliadis \& Wood (1994) and Bl\"ocker (1995), the 0.66-0.68 
M$_\odot$ post-AGB star of NGC 6741 has recently (a few hundreds years ago) exhausted the hydrogen--shell nuclear burning, and is fading 
along the white dwarf cooling sequence. 

To be noticed: the early luminosity decline of the star was very fast (and caused the nebular recombination), but later 
it gradually slowed, so that  
we cannot exclude that, at present, the gas has reached (or even passed) the contraction-expansion equilibrium condition, 
(1/3)$\times$(d(lnL$_*$)/dt)=-2/t$_{\rm kin}$ in the classical Str\"omgren model, thanks to the matter dilution due to expansion. 

New input will come from the nebular physical conditions, radial ionization structure and photo-ionization model.

\section{The nebular parameters}

\subsection{Physical conditions and ionized mass}
Following Sabbadin et al. (2004, and references therein), the radial profile of the physical conditions is given by the zvpc, 
which is independent on the expansion velocity field, since it represents the gas in the plane of the sky, whose motion is tangential. 
For $T_{\rm e}$ we use the classical diagnostic line ratios of ions in p$^2$ and p$^4$ configurations ($\lambda$5007 $\rm\AA\/$/$\lambda$4363 
$\rm\AA\/$ of [O III] and $\lambda$6584 $\rm\AA\/$/$\lambda$5755 $\rm\AA\/$ of [N II]; Aller 1984, Osterbrock 1989), whereas 
$N_{\rm e}$ comes from both diagnostic line ratios of ions in p$^3$ configuration ($\lambda$6717 $\rm\AA\/$/$\lambda$6731 $\rm\AA\/$ of 
[S II] and $\lambda$4711 $\rm\AA\/$/$\lambda$4740 $\rm\AA\/$ of [Ar IV]), and the absolute H$\alpha$ flux distribution.

The main limitation is connected to the NGC 6741 compactness, large stratification of the radiation, and weakness of the [N II] auroral line and 
the [S II] and [Ar IV] doublets: $T_{\rm e}$[N II], $N_{\rm e}$[S II] and $N_{\rm e}$[Ar IV] can be derived only at (or close to) the 
intensity peak of the corresponding emission.

Moreover, the large H$\alpha$ broadening implies a complex deconvolution for instrumental resolution plus thermal motions plus fine-structure 
(see Sect. 2), lowering the accuracy of the F(H$\alpha)_{\rm zvpc}$ and $N_{\rm e}$(H$\alpha$) profiles. Thus, according to Benetti et al. (2003), 
F(H$\alpha)_{\rm zvpc}$ and $N_{\rm e}$(H$\alpha$)
are also obtained from the radial ionization structure relative to O$^{++}$  (Sect. 7.2) and the fair assumption O/H=constant 
across the nebula; at each position:
\begin{equation}
\frac{\rm F(H\alpha)_{zvpc}}{\rm F(\lambda 5007\AA)_{zvpc}}\propto\frac{H}{O}\times f(T_{\rm e})\times {\rm icf(O^{++})},
\end{equation}
with ${\rm icf(O^{++})}$=${\rm \frac{O}{O^{++}}}$=ionization correcting factor. 

For the internal, high-excitation regions ${\rm icf(O^{++})}$ comes from the ionization structure of helium (Seaton 1968, Benetti et al. 2003): 
\begin{equation}
{\rm icf(O^{++})_{\rm inner}} =  1+ \frac{0.3\times He^{++}}{He^+}; 
\end{equation}

for the external, low-excitation layers we adopt: 
\begin{equation}
{\rm icf(O^{++})_{\rm outer}} = 1 + \frac{O^0}{O^{++}} + \frac{O^+}{O^{++}}, 
\end{equation}
which includes the ionization effects produced by the resonant charge-exchange reaction O$^+$ + H$^0$$\getsto$O$^0$ + H$^+$; 
according to Osterbrock (1989) and Stancil et al. (1999), in the outermost nebular regions  H$^0$/H$^+$$\simeq$0.83$\times$(O$^0$/O$^+$).

Moreover (Sabbadin et al. 2004), 

\begin{equation}
N_{\rm e}(H\alpha)\propto \frac{1}{T_{\rm e}^{-0.47}} \times 
(\frac{\rm F(H\alpha)_{zvpc}}{\epsilon_{\rm l} \times {\rm D}})^{1/2},
\end{equation}
where D is the (known) distance, and $\epsilon_{\rm l}$ the (un-known) ``local filling factor'', i. e. the fraction of the local 
volume actually filled by matter with density $N_{\rm e}$. 
 
A comparative analysis gives quite satisfactory results: the $N_{\rm e}$(H$\alpha$) profiles obtained in the two ways 
differ by less than 5$\%$ everywhere, except in the faint, innermost regions, where the discrepancy raises up to 10$\%$. In the following we 
will adopt $N_{\rm e}$(H$\alpha$) given by F($\lambda$5007 $\rm\AA\/$)$_{\rm zvpc}$ and Eqs. (9) to (12).

Fig. 13 shows the resulting radial distribution of the physical conditions in NGC 6741 at PA=15$\degr$ and PA=75$\degr$ (close to the apparent 
minor and major axes, respectively).
 
  \begin{figure*} \centering
   \includegraphics[width=17cm,height=17cm]{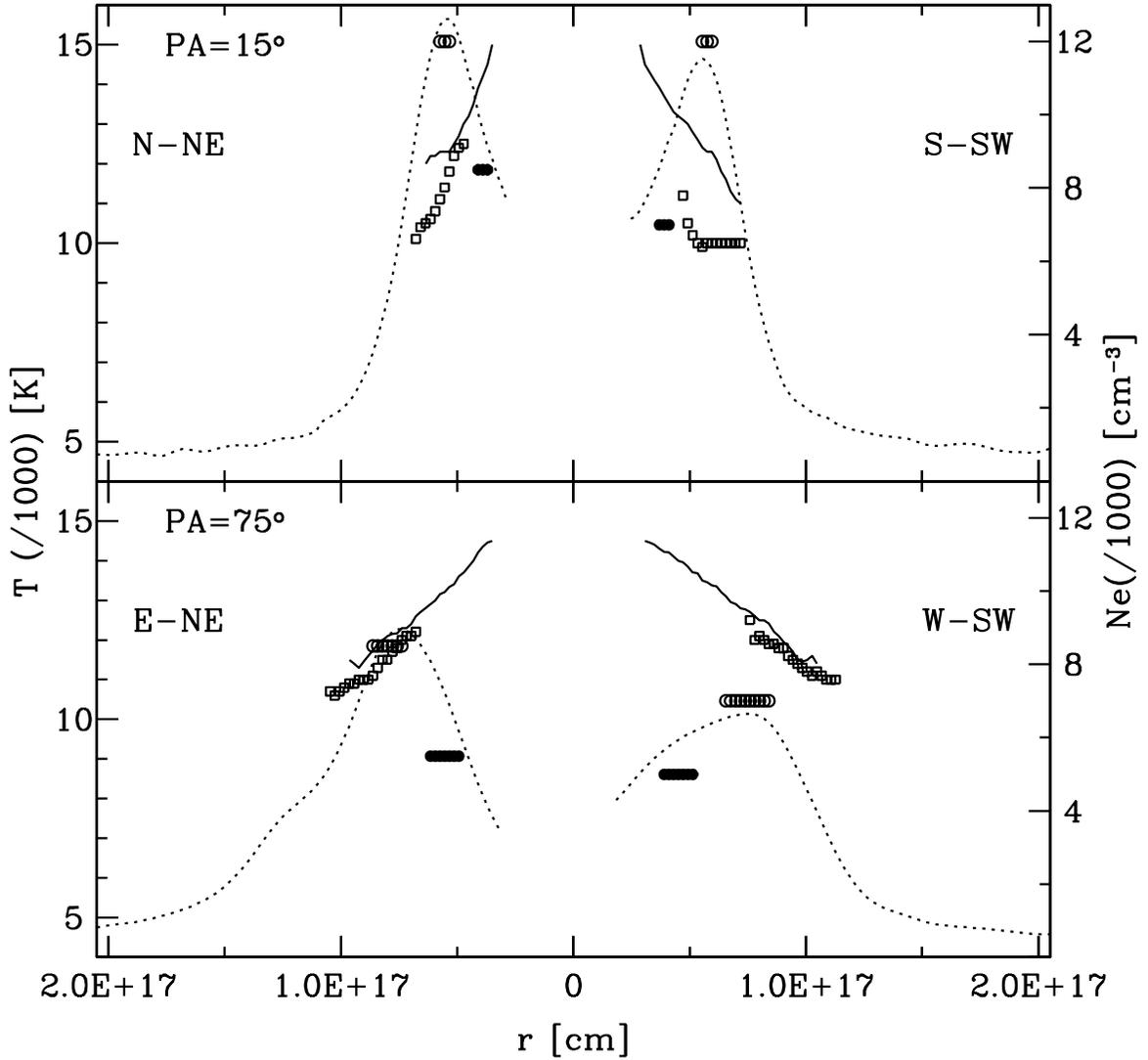}
\caption{Radial distribution of the physical conditions ($T_{\rm e}$ and $N_{\rm e}$) at two selected directions of NGC 6741. Upper panel: the 
zvpc at PA=15$\degr$ (close to the apparent minor axis); lower panel: the zvpc at PA=75$\degr$ (close to the apparent major axis). 
Left ordinate scale:  $T_{\rm e}$[O III] (continuous line) and 
$T_{\rm e}$[N II] (empty squares). Right ordinate scale: $N_{\rm e}$(H$\alpha$) for D=2000 pc (dotted line), $N_{\rm e}$[S II] (empty circles), and 
$N_{\rm e}$[Ar IV] (filled circles).}
   \end{figure*}
The radial trend of $T_{\rm e}$[O III] is common at all PA: $\ge$15\,000 K in the 
weak innermost regions, rapidly decreasing outward down to 12\,000 K in the densest layers, and 
more or less constant furtherout (uncertain). $T_{\rm e}$[N II] is systematically lower than  $T_{\rm e}$[O III]. All this is  in quantitative 
agreement with the results by Hyung \& Aller (1997) and Pottasch et al. (2001). 
 
$N_{\rm e}$ presents a broad, asymmetric bell-shape profile with peaks up to 12\,000  ($\pm$1000) cm$^{-3}$ close to the minor axis, 
and 6000--8000 ($\pm$800) cm$^{-3}$ close to the major axis. 
Note that, in spite of the high density peaks - causing strong collisional de-excitation of the [S II] $\lambda$6717 $\rm\AA\/$ line - 
we have $N_{\rm e}$[S II]$\simeq$$N_{\rm e}$(H$\alpha$), i. e. the local filling factor is $\epsilon_{\rm l}$$\simeq$1, being  
$N_{\rm e}$[S II] $\times \epsilon_{\rm l}^{0.5}\simeq N_{\rm e}$(H$\alpha)$ (Aller 1984, Pottasch 1984, 
Osterbrock 1989).

$N_{\rm e}$[Ar IV] is systematically lower than $N_{\rm e}$[S II] adopting the electron impact excitation rates 
by Keenan et al. (1997), whereas  $N_{\rm e}$[Ar IV]$\simeq N_{\rm e}$[S II] for earlier collisional rates (for example, Aller 1984). 

Concerning the halo, $N_{\rm e}$ peaks at the inner edge ($\simeq$1000 - 1500 cm$^{-3}$), and decreases outwards.

Recent density determinations (mean values integrated over the slit) reported in the literature for NGC 6741 are: 
$N_{\rm e}$(different ions)=6300 cm$^{-3}$ by Hyung \& Aller (1997), $N_{\rm e}$[S II]=6000 cm$^{-3}$,  $N_{\rm e}$[O II]=8500 cm$^{-3}$, 
$N_{\rm e}$[Cl III]=9000 cm$^{-3}$ and $N_{\rm e}$[Ar IV]=5500 cm$^{-3}$  by Pottasch et al. (2001), and $N_{\rm e}$[Cl III]=4470 cm$^{-3}$, 
$N_{\rm e}$[Ar IV]=6610 cm$^{-3}$ and $N_{\rm e}$[52$\mu$m/88$\mu$m]=2880 cm$^{-3}$ by Liu et al. (2001).

The ionized mass of NGC 6741, obtainable from the observed $N_{\rm e}$ spatial distribution, the H$\beta$ flux, and 
the radio flux (Aller 1984, Pottasch 1984, Osterbrock 1989, and Turatto et al. 2002), results to be M$_{\rm ion}$$\simeq$0.06($\pm$0.02) 
M$_\odot$, i. e. much lower than the mass of the external, neutral cocoon, M$_{\rm neutral}$$\simeq$0.20 ($\pm$0.05) M$_\odot$ (see Sect. 4), 
thus confirming the peculiar evolutionary phase of our nebula.

To be noticed: the relatively high $N_{\rm e}$ of the halo implies a quite recent recombination start, no more 
that 200 years ago. In fact, the $N_{\rm e}$ depletion rate for 
recombination is: 
\begin{equation}
{\rm d}N_{\rm e}/{\rm dt} = -\alpha_{\rm B}\times N_{\rm e}\times N({\rm H}^+), 
\end{equation}

with $\alpha_{\rm B}$=effective recombination coefficient.
Assuming $N_{\rm e}\simeq N({\rm H}^+$) and $T_{\rm e}$=12\,000 K, and neglecting the recombination delay 
due to expansion, we obtain:
\begin{equation}
N_{\rm e}(0) = N_{\rm e}({\rm t})/[1 - 8.2\times10^{-6}\times{\rm t}\times N_{\rm e}{\rm (t)}],
\end{equation}
where $N_{\rm e}$(0) is the initial electron density, and $N_{\rm e}$(t) the electron density 
at time t (in years) elapsed from the recombination start. Adopting  $N_{\rm e}$(0)$\simeq$10\,000 cm$^{-3}$ (mean of the density peaks 
in Fig. 13), 
and $N_{\rm e}$(t)$\simeq$600 cm$^{-3}$ (the lowest value in Fig. 13), we have t$\simeq$190 years. A very similar ``recombination age'' is 
inferred from the mean H I density of the circum-nebular matter (Sect. 3). 

All this, combined with the short nebular age ($\simeq$1400 years), fairly agrees with the theoretical evolutionary 
times of a 0.66-0.68 M$_\odot$ post-AGB star. According to  Bl\"ocker (1995 and references therein), in the interval t$_{\rm SW}$ 
(end of the superwind ejection) to  t$_{\rm SW}$+1100yr the hydrogen-burning star crosses horizontaly 
the H-R diagram (at log L$_*$/L$_\odot\simeq$4.00) 
becoming hotter and hotter (up to logT$_*\simeq$5.20 K). At t$_{\rm SW}$+1100yr the nuclear fuelling becomes insufficient, and  
in a century the luminosity  
declines to log L$_*$/L$_\odot\simeq$3.70 (whereas T$_*$  reaches its maximum, logT$_*\simeq$5.30 K). At t$_{\rm SW}$+1200yr the 
hydrogen-shell burning ceases at all, 
the stellar evolution drastically accelerates till t$_{\rm SW}$+1400yr (log L$_*$/L$_\odot\simeq$2.80; logT$_*\simeq$5.20 K), and later 
slows down (for example, at t$_{\rm SW}$+3000yr we have log L$_*$/L$_\odot\simeq$2.40 and  logT$_*\simeq$5.15 K).

Summing up: NGC 6741 is close to (or has even passed) the recombination-reionization transition.

\subsection{Radial ionic and mean chemical abundances, and photo-ionization model}

The zvpc of the different emissions furnishes the detailed radial ionization structure of NGC 6741. Since the blurred H$\alpha$ appearance  
lowers the accuracy of the 
F(H$\alpha)_{\rm zvpc}$ distribution, we adopt $\lambda$5007 $\rm\AA\/$ as reference emission, thus  
inferring the radial ionization structure relative to O$^{++}$ (for details, see Benetti et al. 2003). The $\frac{X^{+i}}{O^{++}}$ profiles 
of NGC 6741 at PA=15$\degr$ and PA=75$\degr$ (close to the apparent minor and major axes, respectively), presented in Fig. 14, 
confirm that the nebula is optically thick at all directions.

Within this scenario, the [O III] rays punching the [N II] skin (Fig. 2) identify a few radial directions along and close to the true 
major axis of the nebula with ``low''  $N_{\rm e}$ peaks ($N_{\rm e}$(peak)$\le$4000 cm$^{-3}$ according to the photo-ionization model 
presented in the course of the section): in these directions (barely visible in our spectral images at PA=75$\degr$, 95$\degr$ and 115$\degr$; 
see Figs. 4 and 10) recombination processes are less efficient, and the gas can be (partially) ionized by the UV stellar flux at larger 
distanced. 

When combined with the results stored in the previous sections, all this indicates that the overall, complex structure of NGC 6741, fully driven by 
the fast evolving central star, can be divided into three distinct zones showing quite different physical characteristics: the internal, 
high-to-medium excitation nebula directly ionized by the stellar radiation ends in a sharp transition zone at medium-to-low excitation 
marking the present edge of the ionization front. It is surrounded by the gas no more reached by the fading UV flux (i.e. the halo), dominated 
by recombination processes. A similar onion-like radial structure (ionization-transition-recombination) is observed in NGC 6565 (Turatto 
et al. 2002).

Though the ionization and thermal structure of NGC 6741 result to be out of equilibrium, the large $N_{\rm e}$ of the ionized gas (peaks 
up to 12\,000 cm$^{-3}$) assures a quick nebular reply to the variable stellar flux. In particular, the delay in the ``transition zone'' 
is short, the recombination time, t$_{\rm rec}$=1/($\alpha_{\rm B} \times N_{\rm e}$), being a few years for hydrogen (even shorter 
for higher ionization species). 

In conclusion: NGC 6741 is nearly in equilibrium. This allows us to neglect (in first approximation) the time dependance, and estimate 
the total chemical abundances of the ionized gas with the classical method valid for ``static'' nebulae. 

\begin{figure*} \centering
\includegraphics[width=18cm,height=20cm]{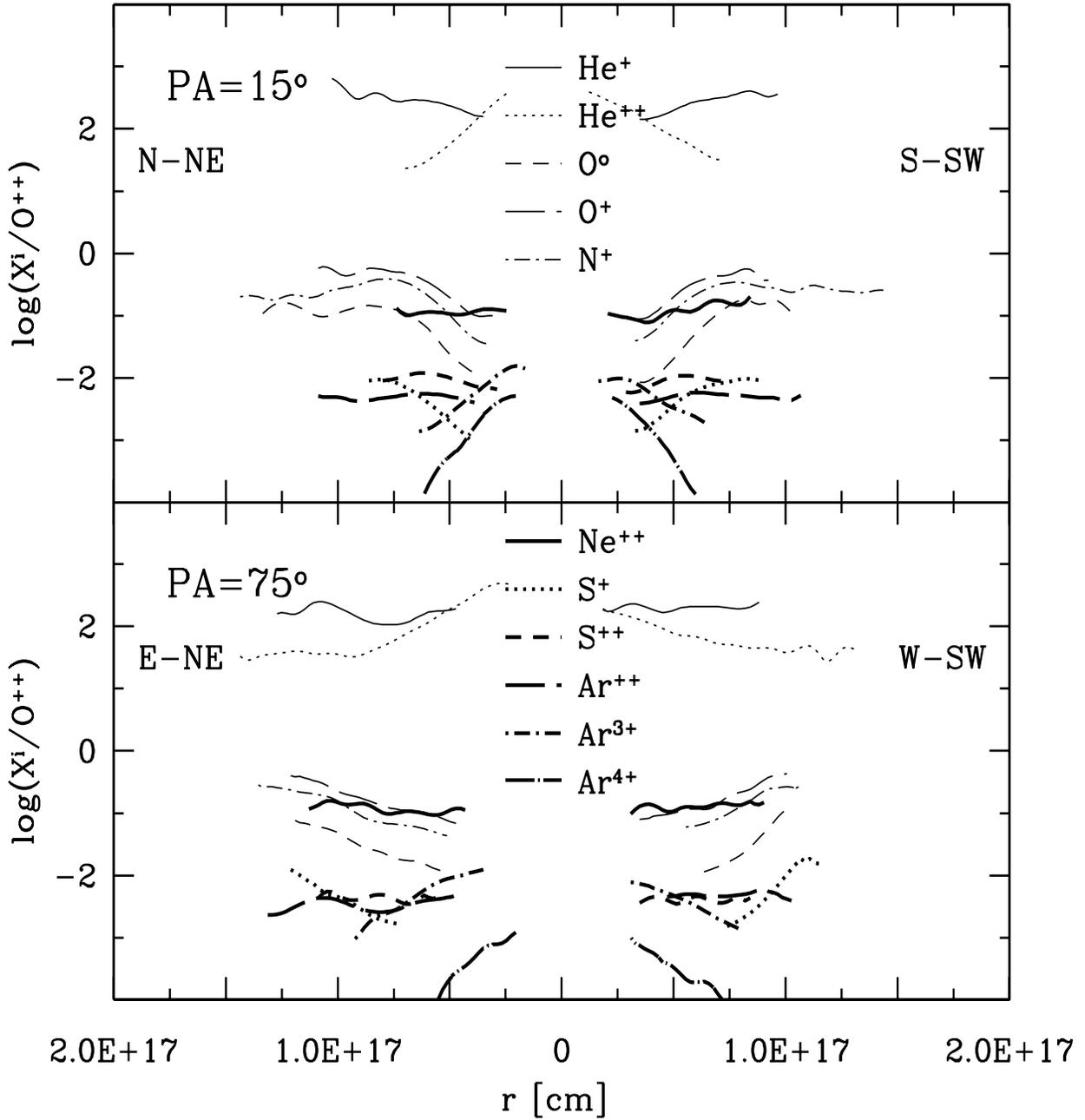} 
\caption{The $\frac{X^{+i}}{O^{++}}$ radial ionization structure of NGC 6741 in the 
zvpc at PA=15$\degr$ (close to the apparent minor axis; upper panel), and PA=75$\degr$ (close to the apparent major axis; lower panel).
Same orientation as Fig. 13.}  
\end{figure*}
We follow a two-step procedure:
\\
{\bf I (broad approach)}: according to the critical analyses by Alexander \& Balick (1997) and Perinotto et al. (1998), the mean ionic 
abundances are obtained from the line fluxes integrated over the spatial profile and the expansion velocity field; later on, we correct
for the unobserved ionic stages by means of the ionization correcting factors, derived both empirically (Barker
1983, 1986), and from interpolation of theoretical nebular models (Shields et al. 1981, Aller \& Czyzak 1983, Aller 1984, Osterbrock 1989). 
The resulting mean chemical abundances are listed in Table 3 (last but one row); the recent estimates reported in the literature are 
also given in the Table. 

\begin{centering}
\begin{table*}
\caption{NGC 6741: chemical abundances (relative to log H=12)}
\begin{tabular}{lcccccccc}
\hline
\\
Reference& He& C& N& O &Ne&S&Cl&Ar\\
\\
\hline
\\
Hyung \& Aller 1997 (icf)&11.01&8.86&8.38&8.73&8.12&6.89&5.43&6.49\\
\\
Hyung \& Aller 1997 (model B)&11.04&8.90&8.15&8.65&8.00&6.76&5.34&6.54\\
\\
Pottasch et al. 2001&11.04&8.81&8.45&8.82&8.26&7.04&-&6.69\\
\\
Pottasch et al. 2002&11.04&8.56&8.26&8.65&8.18&6.90&5.26&6.51\\
\\
Perinotto et al. 2004b&11.08&- &8.31&8.70&8.16&6.70&- &6.54\\
\\
This paper (icf) &11.04&-&8.20&8.66&8.08&6.78&- &6.54\\
\\
This paper (CLOUDY) &11.04&-&8.28&8.66&8.08&6.90&- &6.54\\
\\ 
\hline
\end{tabular}
\end{table*}
\end{centering}  

{\bf II (refining)}: we apply the photo-ionization code CLOUDY (Ferland et al. 1998) to a model-nebula characterized by the same 
distance, gas distribution, mean chemical composition and exciting star parameters of NGC 6741, and then combine the model-line profiles 
(convolved for a seeing+guiding of 0.60$\arcsec$) with the line profiles observed in the true nebula.  

Let us focus on the zvpc of the S-SW sector at PA=15$\degr$ (close to the apparent minor axis of NGC 6741). The input parameters of the model-nebula 
are given in Table 4, whereas Fig. 15 shows the physical conditions in the model-nebula (to be compared with Fig. 13), and the absolute 
radial flux distribution of the main emissions in both the model-nebula and NGC 6741. 

At first glance, Fig. 15 provides a satisfactory model-nebula vs. true-nebula agreement for the internal, high-to-medium excitation regions and the 
transition zone. Furtherout, the weak recombining halo appears in NGC 6741, whereas emissions are absent at all in the static model-nebula. 
The same is observed at the other PA. 

When entering in more detail, some disturbing drifts arise in Fig. 15. Two minor discrepacies are:
\begin{description}
\item[(a)]  uncertain flux distribution of  
[Ne III] in NGC 6741 (ascribable to the edge location of $\lambda$3968 $\rm\AA\/$ in the original NTT+EMMI frames), 
\item[(b)] comparable model-nebula vs. true-nebula flux-shift for [S II] and [S III], indicating a slight underestimate of 
sulphur abundance 
in Table 3 (last but one row), and in Table 4 (model-nebula). The same occurs for [N II]. The improved chemical abundances of NGC 6741 are 
reported in the last row of Table 3. 
\end{description}
The two major problems in Fig. 15 concern:
\begin{description}
\item[(1)] the ionization  structure of helium: He I $\lambda$5876 $\rm\AA\/$ and He II $\lambda$4686 $\rm\AA\/$ are weaker and 
stronger, respectively, in the model-nebula than in the true-nebula;
\item[(2)] the peak emission and whole flux distribution of the highest excitation ions (He II, [Ar IV] and [Ar V]) 
tend to be systematically closer to the central star in the true-nebula than in the model-nebula.
\end{description}

Ad hoc manipulation of the model-nebula input data lowers - or even cancels - the foregoing problems; for example, we can act on the ionizing 
UV distribution, given the strong T$_*$ dependance of both the ionic structure of helium and the radial profile of the 
highest-excitation emissions. 
In fact, the only decrease by 17$\%$ of the stellar temperature (i.e. 
for T$_*\simeq$150\,000 K, which is within the observational error box given in Sect. 6) simultaneously relieves discrepancies (1) and (2). 
They almost disappear with a further, modest increase (by 20$\%$) of the matter density in the innermost nebular layers.

Unfortunately, the large number of involved parameters (density profile, chemical abundances, dust, temperature and luminosity of the star) and 
assumptions (distance, black-body distribution, seeing convolution), added to the nebular compactness and peculiar evolutionary phase, 
greatly complicate the affair, whose definitive solution needs detailed observations at even higher spatial and spectral 
resolutions, and is beyond the aims of the paper (as well beyond the patience of the authors).

\begin{figure*} \centering
\includegraphics[width=18cm,height=21cm]{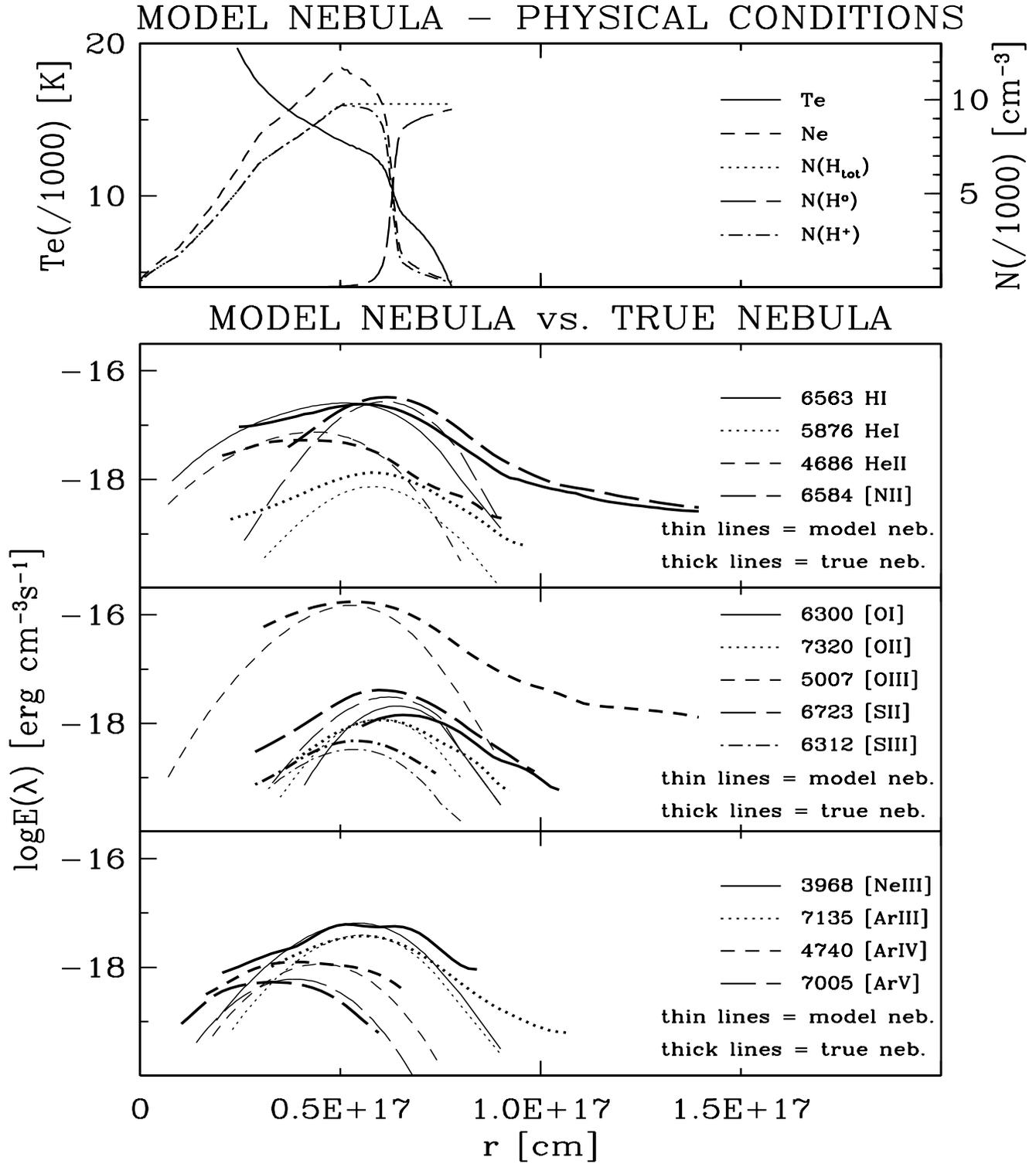} 
\caption{Model-nebula (CLOUDY) vs. true-nebula (NGC 6741, PA=15$\degr$, S-SW sector): physical conditions and 
radial ionization structure.   
Top panel: physical conditions in the model-nebula (to be compared with Fig. 13). Second panel: absolute flux distribution of 
H I $\lambda$6563 $\rm\AA\/$, [N II] $\lambda$6584 $\rm\AA\/$, and the ionic sequence of helium (He I $\lambda$5876 $\rm\AA\/$ and He II 
$\lambda$4686 $\rm\AA\/$) in the model-nebula (thin symbols) and the true-nebula (thick symbols). Third panel: 
absolute flux distribution for the ionic sequences of oxygen ([O I] $\lambda$6300 $\rm\AA\/$, 
[O II] $\lambda$7320 $\rm\AA\/$, and [O III] $\lambda$5007 $\rm\AA\/$) and sulphur ([S II] $\lambda$6717+6731 $\rm\AA\/$, and [S III] 
$\lambda$6312 $\rm\AA\/$); same symbols as in the second panel. Bottom panel: absolute flux distribution for [Ne III] $\lambda$3968 $\rm\AA\/$, 
and the ionic sequence of argon ([Ar III] $\lambda$7135 $\rm\AA\/$, [Ar IV] $\lambda$4740 $\rm\AA\/$, and [Ar V] $\lambda$7005 $\rm\AA\/$); 
same symbols as in the second and third panels.}  
\end{figure*}

\begin{table}
\caption{Input parameters for the model nebula (CLOUDY)}
\begin{tabular}{ll}
\hline
\\
Radial density profile &  ionized gas = Fig.~13; cfr. Sect. 7.1\\
&neutral gas = cfr. Sects. 3 and 4\\
\\
Chemical abundances:   & \\
~~ C, Cl, K, Ca, Mg, Si & Hyung \& Aller (1997) model B\\
~~ He, N, O, Ne, S, Ar & Table 3 (last but one row)\\
~~ other elements      & PN (CLOUDY default)\\
&\\
Dust                   & PN (CLOUDY default)\\
&\\
Local filling factor         & 1.0 \\
&\\
&blackbody distribution\\
Exciting star   & T$_*$=170\,000 K \\
& log L$_*$/L$_\odot$= 2.75\\
&\\
Distance& 2.0 kpc\\
\\
\hline
\end{tabular}
\end{table}

\section{The 3-D morpho-kinematical structure}
 
\begin{figure*} 
\centering
\includegraphics[width=15cm]{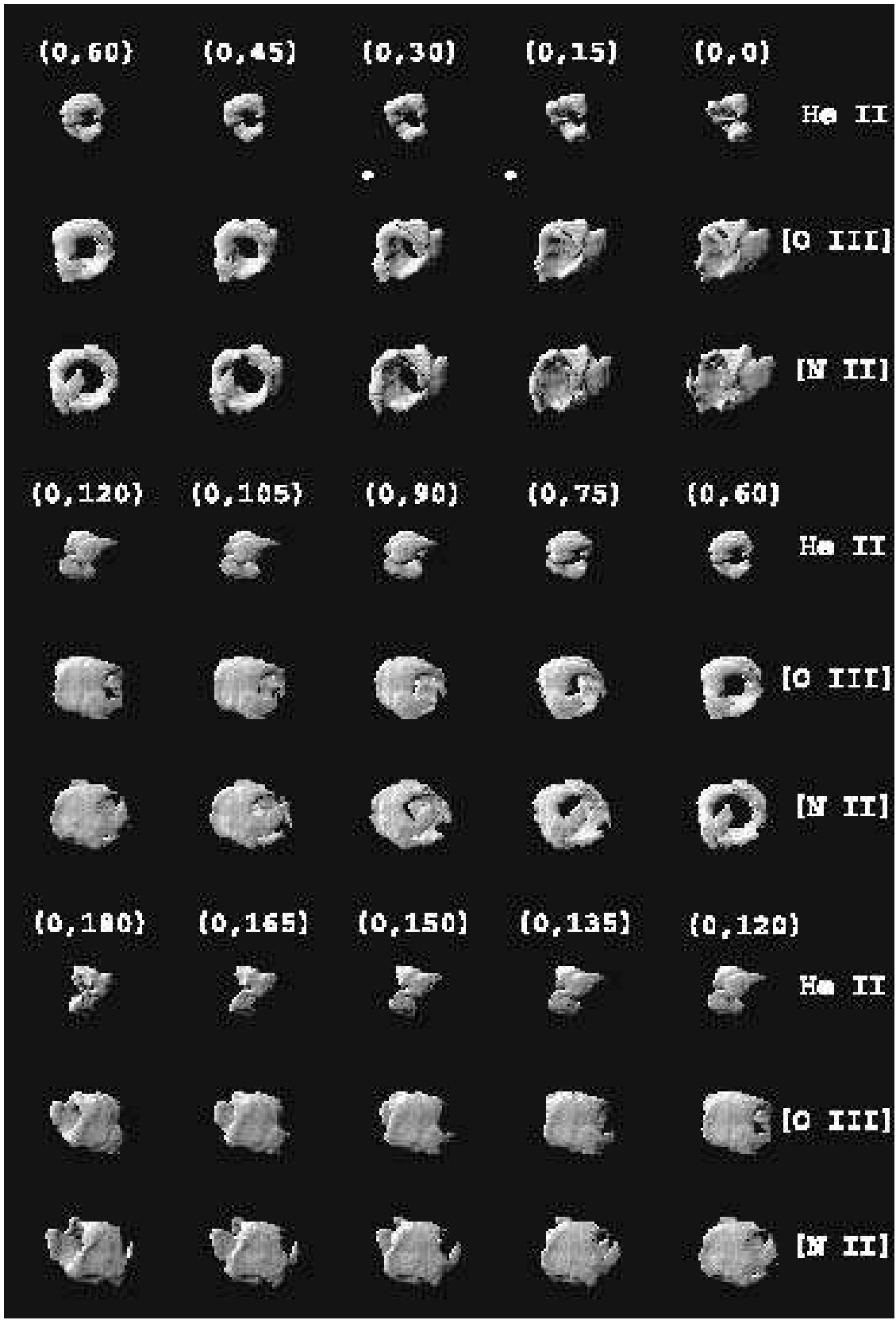} 
\caption{Stereoscopic structure of NGC~6741 for a rotation around the 
North-South axis centered on the exciting star. Opaque reconstruction at high flux-cut for $\lambda$4686 $\rm\AA$
of He II, $\lambda$5007 $\rm\AA$ of [O III], and $\lambda$6584 $\rm\AA$ of [N II], as seen from thirteen directions separated by 
15$\degr$. The line of view
is given by ($\theta,\psi$), where $\theta$ is the zenith angle and $\psi$ the
azimuthal angle. Each horizontal pair represents a ``direct''
stereoscopic pair (i. e. parallel eyes), and the whole figure  provides twelve 3-D views of the nebula in as
many directions, covering a straight angle (for details, see Ragazzoni et al. 2001). The (0,0) images represent the rebuilt-nebula as 
seen from Earth (North is up and East to the left). The complete series of nebular movies is shown in the electronic version of the paper 
(on-line data).}  
\end{figure*}
\begin{figure*} 
\centering
\includegraphics[width=15cm]{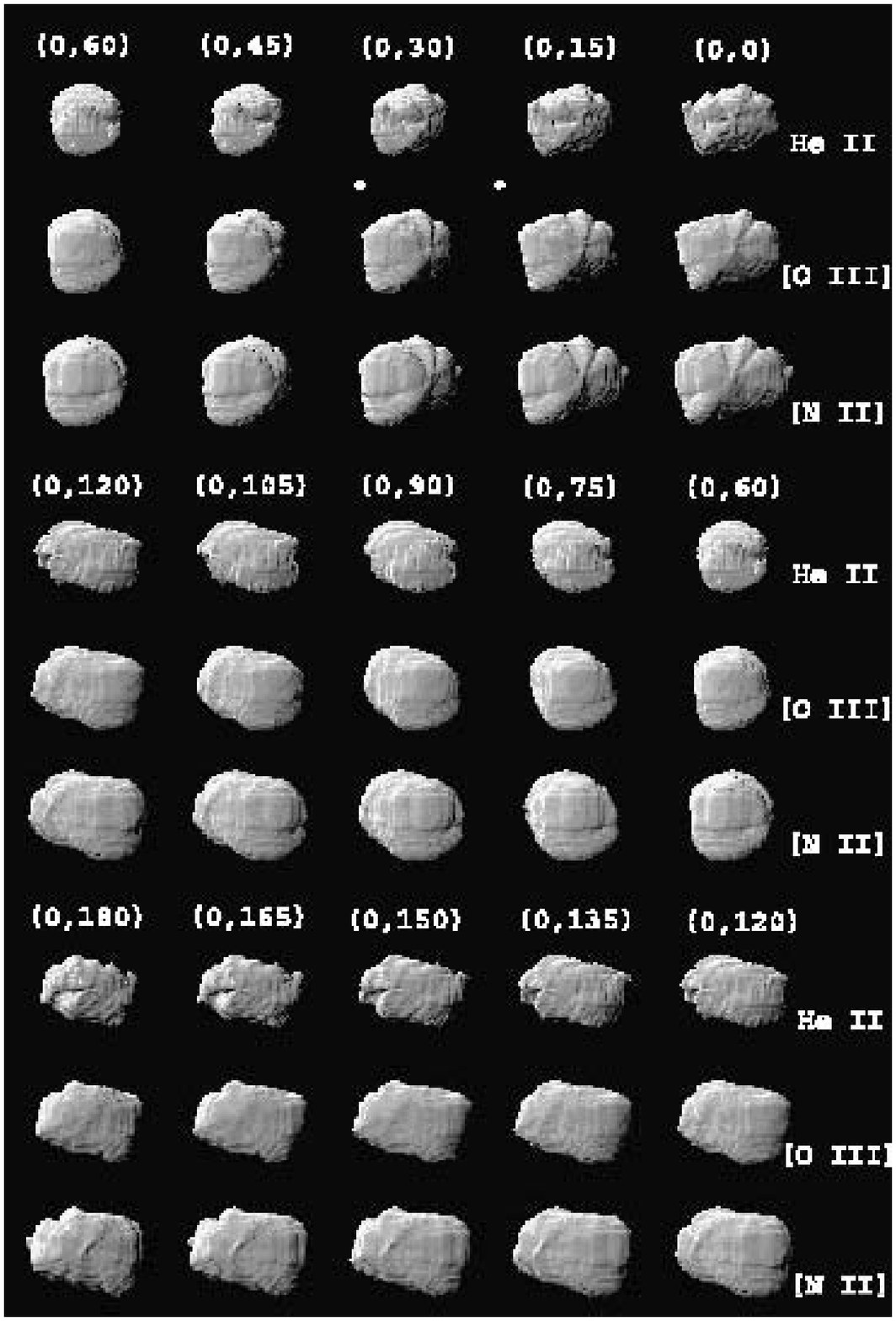} 
\caption{Same as Fig. 16, but at low flux-cut. 
}  
\end{figure*}
According to the reconstruction method introduced by Sabbadin et al. (1985, 1987), and deepen by Sabbadin et al. (2000a, b), 
we select $\lambda$4686 $\rm\AA$ 
of He II, $\lambda$5007 $\rm\AA\/$ of [O III] and $\lambda$6584 $\rm\AA\/$ of [N II]
as markers of the high, medium and low-ionization regions of NGC 6741, respectively. The spectral images are:
\\
(a)  de-convolved for seeing, spectral resolution and thermal motions (also fine-structure is taken into account for 
the recombination line of He II),
\\
(b) de-projected through the relation  $V_{\rm exp}$(km s$^{-1}$)= 13$\times$R$\arcsec$ (see Sect. 3),
\\
(c) assembled by means of the 3-D rendering procedure described by Ragazzoni et al. (2001).

Here we present a limited number of frames, corresponding to a partial rotation around 
the North--South axis centered on the exciting star (i.e. almost perpendicular to the major axis).  
The complete series of nebular movies is available:

- in the electronic version of the paper (on-line data),
  
- at  {\bf http://web.pd.astro.it/sabbadin}, the WEB site dedicated to the 3-D structure of expanding nebulae, 

- at {\bf http://www.edpsciences.org}. 
 
Figs. 16 and 17 show the opaque reconstruction of NGC 6741 in He II, [O III] and [N II] at high and low flux-cuts, respectively, for a 
rotation of 180$\degr$ through the first Euler angle. The 
(0,0) images correspond to the Earth--nebula direction (North is up and East to the left). 
An indication of the $N_{\rm e}$ limit in Figs. 16 and 17 can be obtained from the corresponding [O III] absolute flux-cut: log 
E($\lambda$5007 $\rm\AA$)= -15.95 and -16.70 erg s$^{-1}$ cm$^{-3}$ (high and low cuts, respectively). 
Since O/H=4.6$\times$10$^{-4}$ (Table 3),  ${\rm icf(O^{++})}$$\simeq$1.1, 
and $N_{\rm e}\simeq$1.15$\times$N(H$^+$), they correspond to 
$N_{\rm e}$(high cut)$\simeq$9000 
cm$^{-3}$ for Fig. 16, and $N_{\rm e}$(low cut)$\simeq$4000 cm$^{-3}$ for Fig. 17 (assuming $T_{\rm e}$=12\,000 K and 
$\epsilon_{\rm l}$=1). 

The projection of NGC 6741 
for a rotation around the N--S direction (almost perpendicular to the major axis) is presented in Fig. 18 (multi-color images only in the ``free'' 
electronic version of the paper, for lack of funds), providing 
a representative sample of the nebular appearance when changing the line of view. The left panel, (0,0),  corresponds to NGC 6741 as 
seen from Earth (North is up and East to the left), to be compared with Fig. 1 and the HST image by 
Hajian \&  Terzian at {\bf http://ad.usno.navy.mil/pne/gallery.html}. 

The overall characteristics of NGC 6741 (a hot and faint star at the centre of a dense, inhomogeneous, equatorial torus merging into a closed, 
ellipsoidal structure embedded into a large cocoon of almost neutral matter) closely resemble NGC 6565, a compact, recombining 
PN projected pole-on (Turatto el al. 2002), and show remarkable affinities with NGC 7027, the PNe prototype (Cox et al. 2002, Bains et al. 
2003). 
Very likely, all three  objects represent ``middle-aged'' snapshots of a  massive, fast evolving post-AGB star. 

\begin{figure*} \centering
\includegraphics[width=17.9cm]{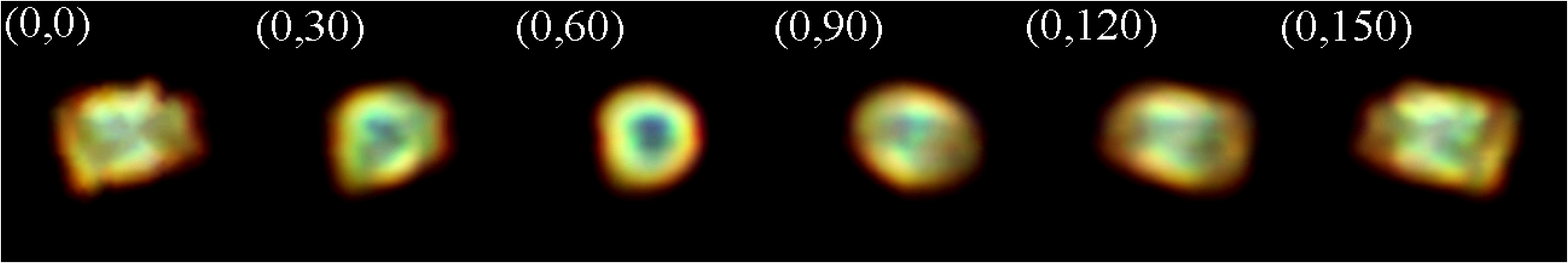} 
\caption{Optical appearance of NGC 6741  (multi-color in the electronic version of the paper; blue=He II, green=[O III], red=[N II]) for a rotation 
through the N--S axis centered on 
the exciting star. The left panel, (0,0), corresponds to the re-built 
nebula as seen from the Earth (North is up and East to the left). Recall that 
projection($\theta,\psi$)=projection($\theta\pm180\degr,\psi\pm180\degr$). The complete (multi-color) movie is shown in the electronic version 
(on-line data).}  
\end{figure*}

\section{Discussion} 

In spite of the nebula compactness, we have quite successfully applied the tomographic and 3-D analyses
to NGC 6741, covered at high spatial and spectral resolutions with ESO NTT+EMMI. 

NGC 6741 is a young (age$\simeq$ 1400 years) PN at a distance of 2000 pc; it consists of a high-excitation, almost-prolate ellipsoid 
(0.039 pc x 0.022 pc x 0.019 pc; major, intermediate 
and minor semi-axes, respectively) denser at the equator than at the poles  
($N_{\rm e}$ peaks up to 12\,000 and 7000 cm$^{-3}$, respectively), surrounded by a thin skin at low-excitation (the transition zone, 
corresponding to the present ionization front), and embedded 
into an extended, spherical (radius$\simeq$0.080 pc), almost neutral recombining halo containing a large fraction of the nebular mass 
(M$_{\rm halo}\simeq$0.20 M$_\odot$ versus M$_{\rm ion}\simeq$0.06 M$_\odot$).

The complex ionization structure of NGC 6741 is fully driven by the fast evolution of the central star, a  massive (0.66-0.68 M$_\odot$), 
hot (log T$_*\simeq$5.23 K) and faint (log L$_*$/L$_\odot\simeq$2.75) post--AGB star fading towards the white-dwarf region (after exhaustion 
of the hydrogen--shell nuclear burning).

The ionized gas of NGC 6741 follows the general expansion law $V_{\rm exp}$(km s$^{-1}$)=13$\times$R$\arcsec$, except the innermost, 
highest-excitation layers, showing clear evidences of deceleration. This peculiar kinematical behaviour can be qualitatively ascribed to the 
luminosity drop of the central star (Sch\"onberner et al. 1997, Steffen et al. 1998, Marigo et al. 2001, Perinotto et al. 2004a): in this 
evolutionary phase 
the stellar mass-loss quickly decreases, and the falling hot-bubble's pressure no more balances the 
pressure of the ionized gas, so that the contact discontinuity and the adjacent, high-excitation nebular layers accelerate inwards. 

The argument is poorly known, and deserves more attention, the kinematical properties of the innermost 
layers being an excellent diagnostic of the star-nebula interaction. A search for more ``decelerated'' candidates, 
a high-resolution spectroscopic survey in ionic species at large IP, and the detailed comparison with the current hydro-dynamical simulations 
are highly desired.  

Note that the opposite situation (i. e. acceleration) is expected for the external, recombining layers of NGC 6741, whose de-pressuring gives 
rise to Rayleigh-Taylor instability at the ionization edge. Though the knotty and filamentary [N II] nebular appearance (Fig. 1, lower panel) 
qualitatively agrees with this scenario, its confirmation needs very-high spectral resolution (R$\ge$100\,000) echellograms detecting the 
increase of expansion velocity (and turbulence!) at the ionization edge. 

A further interesting point concerns the wide radial matter distribution of NGC 6741 (much wider than the $N_{\rm e}$ profile given in 
Fig. 13, due to the presence of the dense, almost-neutral recombining halo). This indicates a modest contribution of wind interaction to the 
nebular shaping 
(it essentially supports the innermost layers, avoiding gas infall), whereas the lion's share is taken by ionization: the thermal pressure 
of the ionized gas accelerates the matter outwards and decelerates inwards, creating the density and velocity distributions observed in NGC 
6741 (more general comments are towards the end of the section). 

Though the precise radial profile of the matter remains unknown - a large fraction of NGC 6741 being neutral -, a more sign provides that the 
outer radius 
of the main nebula extends well beyond the present ionization edge: the absence of any extended, diffuse, faint and roundish 
AGB-halo emission in the Palomar Sky Survey plates, deep imaging by Schwarz et al. (1992), and HST frames.  This is confirmed by our spectra: 
no kinematical signature of an  AGB-halo (i. e. external, faint, un-tilted and un-split emission) appears. 
Since the low-density gas of an AGB-halo (usually $N_{\rm e}$=50 to 200 cm$^{-3}$; Chu et al. 1987; Corradi et al. 2003) is little affected by 
recombination 
(from Eq. (14), $N_{\rm e}$(0)=50 to 200 cm$^{-3}$ implies $N_{\rm e}$(200yr)=47 to 162 cm$^{-3}$), we infer that:
\begin{description}
\item[-] NGC 6741 never became optically thin to the UV stellar radiation,
\item[-] the total nebular mass, M$_{\rm tot}$, is larger than 0.40 M$_\odot$.
\end{description}
Let us focus on recombination, which appears as the main characteristic (and peculiarity) of our nebula. 
As outlined in the previous sections, NGC 6741 is the second PN - of 
the six so far analysed with the 3-D methodology - in a 
deep recombination phase (the first being NGC 6565, Turatto et al. 2002). Moreover, according to Benetti et al. (2003), a third nebula, 
NGC 6818, is at the very beginning of recombination. Such an apparent excess of recombining PNe is a mere consequence of our target 
selection criteria (in particular, high surface brightness and richness of ionic species), favouring massive nebulae powered by fast 
evolving central stars.

Thanks to the unprecedented accuracy achieved by tomography and 3-D recovery on the manifold facets of the PN research, we have identified a 
number of observational evidences supporting the recombination hypothesis for NGC 6741; they are:  
\\
(1) nebula much brighter than the star (Sects. 1 and 6), 
\\
(2) presence of absorbing knots (Sect. 1), 
\\
(3) co-existence of ionic species in a large IP range (Sects. 1, 3 and 7.2), 
\\
(4) kinematics of the halo (Sect. 3), 
\\
(5) high density of the almost-neutral halo (Sects. 3 and 4), 
\\
(6) c(${\rm H}\beta)$ variable over the nebula (Sect. 4), 
\\
(7) high temperature and low luminosity of the central star (Sect. 6),
\\
(8) T$_{\rm Z}$H I$>$T$_{\rm Z}$He II (Sect. 6).

A minimal, but effective, recombination standard for the whole PN class - in practice, a summary of points (1) and (7) - is given by:
\begin{equation}
{\rm m(V_*)_{obs} + log F(H\beta)_{obs}} - 1.1\times c({\rm H}\beta)>5.0, 
\end{equation}
where F${\rm (H\beta)_{obs}}$ is in erg cm$^{-2}$ s$^{-1}$.

A quick look at the Strasbourg--ESO Catalogue of Galactic PNe (Acker et al. 1992) allowed us to select some fifty targets satisfying 
Eq. (15). Half of them (e. g. NGC 2440, Hb 4, NGC 6565, NGC 6537, NGC 6620, NGC 6741, NGC 6886, NGC 6881, NGC 7027, Hu 1-2, NGC 6302, and 
NGC 6563) are compact 
(sometimes with butterfly-like extentions), at high surface brightness, and represent young recombining PNe ejected and 
excited by a massive (M$_*$$\ge$0.64 M$_\odot$) fast evolving star beyong the turn-around point in the H--R diagram. The remaining candidates 
are mean-to-low surface brightness, extended (often bipolar) nebulae (e. g. NGC 6445, 
NGC 6439, A 55, NGC 6772, A 53, NGC 6818, NGC 6894, NGC 7048, NGC 2438, NGC 2818, and IC 4406), 
corresponding to aged, evolved PNe, optically thick (at least in some directions) to the fading UV radiation of a moderately-massive (0.60 to 
0.64  M$_\odot$) post-AGB star in the white dwarf cooling sequence.
 
Further supports to the ``massive'' nature for the central stars of PNe satisfying Eq. (15) are:
\begin{description}
\item [-] most targets belong to the morphological class B of Greig (1971, 1972), 
\item [-] a large fraction consists of Type-I PNe (Peimbert 1978; Peimbert \& Torres--Peimbert 1983; Torres--Peimbert \& Peimbert, 
1983;  Peimbert et al. 1995), 
\item [-] the H$_2$ emission at 2.12 $\mu$m is commonly detected (Kastner et al. 1996; Natta \& Hollenbach 1998; Bohigas 2001). Note that, 
though NGC 6741 is not listed among the H$_2$-emitters, a deep search promises fruitful results.
\end{description}

The general rule for recombining PNe is that, the higher the surface brightness of the nebula, the larger is the mass of the star. 

Summing up: recombination represents a not un-common evolutionary phase in PNe; it becomes unescapable for massive nebulae ejected and 
excited by a massive, fast-evolving post-AGB star (also see Tylenda 1986, Szczerba 1990, Phillips 2000, Marigo et al. 2001, Turatto et al. 
2002, Benetti et al. 2003, and Perinotto et al. 2004a).

Let us close this verbose paper with some general considerations on the kinematics of the whole PN class. The comparative analysis performed in 
Sect. 3 - rejecting the kinematical results based on spherical symmetry assumption and/or emission profiles of recombination lines integrated 
along the slit - enhances the rarity of real nebulae showing a ``U'-shaped expansion velocity field. 

We stress that: 
\begin{description}
\item[(a)] Wilson's law ($V_{\rm exp}$ $\propto$ radius) is the general rule for PNe (Wilson 1950, Weedman 1968, Sabbadin et al. 2004 and references 
therein);
\item[(b)] at present, only BD+30$\degr$3639 and NGC 40 show clear observational 
evidences of a soft acceleration in the innermost layers (Bryce \& Mellema 1999; Sabbadin et al. 2000a). Both objects: (I) are 
very-low excitation PNe (I([O III] $\lambda$5007 $\rm\AA\/$)$<$ I(H$\beta$)), (II) exhibit an extremely sharp radial matter profile (shell 
thickness $\Delta$R/R$\le$0.25), and (III) are powered by a ``cold'' and bright central star of late-WR spectral type. According to Iben et al. 
(1983), Bl\"ocker (1995, 2001), and Herwig et al. (1999), they represent ``born-again'' PNe (i. e. the hydrogen-deficient star suffers a 
late thermal pulse during the motion towards the white-dwarf region. Due to the pulse-induced convection, hydrogen 
mixes and burns on a convective turn-over time scale); 
\item[(c)] three more ``born-again'' candidates - A 30, A 58 and A 78 - are faint, extended PNe presenting a central, complex structure of fast, 
hydrogen-deficient knots (Jacoby 1979, Pollacco et al. 1992). 
\end{description}
All this calls for severe constraints to the 
evolutive parameters responsible of PNe shape and shaping (in particular, wind interaction). In fact, according to the detailed 
radiation-hydrodynamics simulations (Perinotto et al. 2004a, and references therein), a PN is the result of the synergistic effects of ionization 
and fast wind on the gas ejected during the 
superwind phase, generating a typical double-shell structure (inner ``rim'' + outer ``shell'') characterized by a ``U'' or better ``V''-shaped 
expansion velocity field. The double-shell structure may only be destroyed either by recombination of the ``shell''  when the central star 
fades, or by overtaking of the ``shell'' by the ``rim''. In both cases a single-shell configuration emerges, whose expansion velocity field is 
simple and increases always steadily inwards. 

Wind interaction being the main responsible of the ``V''-shaped expansion profile in 
double-shell model-PNe, as well of the increasing inwards velocity field in single-shell model-PNe, we infer that {\bf all current 
radiation-hydrodynamics 
simulations tend to overestimate the dynamical effects of fast stellar wind on the nebular gas} 
(we suspect that post-AGB mass-loss rates adopted by the theoretical simulations are systematically too high). 
Though the same suspect arose in the study of NGC 6565 
(Turatto et al. 2002), NGC 6818 (Benetti et al. 2003) and NGC 7009 (Sabbadin et al. 2004), a quantitative answer (also including: (a) the 
temporal evolution of mass-loss in the superwind phase, Sch\"onberner et al. 2005, and (b) the binarity and the possible role of magnetic 
fields, Garcia-Segura \& Lopez 2000, Blackman et al. 2001) will come from the detailed  
analysis performed on a representative sample of PNe in both hemispheres, covered with ESO NTT+EMMI and TNG+SARG (work in preparation).  

\section{Conclusions}
In our opinion, PN observers should cancel the word ``average'' (and respective synonyms) from their scientific vocabulary: no more average 
line fluxes, integrated  
spectral profiles, mean electron temperature, overall electron density, and so on. In fact, to bridge the historical gap between 
theory and practice (Aller 1994), we need detailed, point-to-point flux and velocity measurements in a wide ionization range, combined with a 
straightforward, versatile methodology of analysis. To this end, we have applied tomography and 3-D recovery (Sabbadin et al. 
2004 and references therein) to long-slit 
echellograms of the compact, bright PN NGC 6741, covered at nine PA with ESO NTT+EMMI.

Our long-lasting journey started with the gas kinematics, penetrated into the galactic and circum-nebular absorptions, and explored the 
nebular distance, mass and age. Later on, we dealt with the stellar properties, pushed through the nebular physical conditions, ionization 
structure and photo-ionization model, went over the stumbling block of image de-projection, and, at last, flew about the multi-color and opaque 
reconstructions of the nebula. 

The wealth and accuracy of the results here presented: 
\begin{description}
\item[-] confirm the unique capacity of tomography and 3-D recovery in extracting the huge amount of physical 
information stored in high-resolution spectroscopy (a tacit invitation, extended to the worldwide researchers),
\item[-] stress once more that properly 
squeezed echellograms do represent a peerless tool for deeping the kinematics, physical conditions, ionization 
structure and evolution of all classes of expanding nebulae (PNe, nova and supernova remnants, shells around Population I Wolf-Rayet stars, 
nebulae ejected by symbiotic stars, bubbles surrounding early spectral-type Main Sequence stars etc.).
\end{description}
\begin{acknowledgements} We wish to thank the support staff of the NTT (in particular, Olivier Hainaut) for the excellent assistance during 
the observations.
\end{acknowledgements}


\end{document}